\documentclass[%
 reprint,
 amsmath,amssymb,
 aps,
]{revtex4-2}

\pdfoutput=1
\usepackage{graphicx}
\usepackage{dcolumn}
\usepackage{bm}
\usepackage{hyperref}
\usepackage{multirow}

\usepackage{amsmath}
\usepackage[font=small,labelfont=bf]{caption}
\usepackage[skip=10pt]{parskip}
\usepackage{ragged2e} 
\usepackage{appendix}

\begin{document}
\preprint{APS/123-QED}

\title{An NV$^-$ center in magnesium oxide as a spin qubit for hybrid quantum technologies}
\author{Vrindaa Somjit}
\email{vsomjit@anl.gov}
\affiliation{%
 Materials Science Division, Argonne National Laboratory, Lemont, IL 60439, USA
}%
\author{Joel Davidsson}%
\email{joel.davidsson@liu.se}
\affiliation{%
Department of Physics, Chemistry and Biology, Link\"oping University, SE-58183, Link\"oping, Sweden
}%
\author{Yu Jin}
\email{jinyuchem@uchicago.edu}
\affiliation{
Pritzker School of Molecular Engineering and Department of Chemistry, University of Chicago, Chicago, IL 60637, USA
}%
\author{Giulia Galli}
\email{gagalli@uchicago.edu}
\affiliation{Pritzker School of Molecular Engineering and Department of Chemistry, University of Chicago, Chicago, IL 60637, USA}
\affiliation{Materials Science Division, Argonne National Laboratory, Lemont, IL 60439, USA}

\date{November 19, 2024}

\begin{abstract}
Recent predictions suggest that oxides, such as MgO and CaO, could serve as hosts of spin defects with long coherence times and thus be promising materials for quantum applications. However, in most cases specific defects have not yet been identified. Here, by using a high-throughput first-principles framework and advanced electronic structure methods, we identify a negatively-charged complex between a nitrogen interstitial and a magnesium vacancy in MgO with favorable electronic and optical properties for hybrid quantum technologies. We show that this NV$^-$ center has stable triplet ground and excited states, with singlet shelving states enabling optical initialization and spin-dependent readout. We predict several properties, including absorption, emission, and zero-phonon line energies, as well as zero-field splitting tensor, and hyperfine interaction parameters, which can aid in the experimental identification of this defect. Our calculations show that due to a strong pseudo-Jahn Teller effect and low frequency phonon modes, the NV$^-$ center in MgO is subject to a substantial vibronic coupling. We discuss design strategies to reduce such coupling and increase the Debye-Waller factor, including the effect of strain and the localization of the defect states. We propose that the favorable properties of the NV$^-$ defect, along with the technological maturity of MgO, could enable hybrid classical-quantum applications, such as spintronic quantum sensors and single qubit gates. 

\end{abstract}

\maketitle


\section{Introduction}
The nitrogen-vacancy center (NV$^-$) in diamond \cite{rodgers2021materials} 
and the divacancy (VV$^0$) in silicon carbide (SiC) \cite{castelletto2020silicon} are prototypical spin defects for quantum technologies, including sensing, communication, and computing. A key requirement for these applications is the long coherence time of the spin defect as well as the scalability and integration of the host into current microelectronic platforms. However, the coherence time of defects in diamond and SiC is intrinsically limited by the presence of non-zero nuclear spin isotopes (namely $^{13}$C and $^{29}$Si, respectively). Additionally, achieving high-yield, scalable fabrication of diamond devices is challenging due to difficulties in etching and polishing diamond \cite{de2021materials}, and the synthesis of high-quality thin films of a single phase of SiC is a challenge due to its many nearly-degenerate polytypes \cite{harmon2022designing}. Therefore, the identification of new host materials for spin defects is an active field of research. 

Recent theoretical work based on cluster correlation expansion simulations predicted that several simple oxide hosts, including calcium oxide (CaO) and magnesium oxide (MgO), could possess long coherence times for spin defects \cite{kanai2022generalized}. The work assumed a model spin defect, i.e. a single electron in a dilute nuclear spin bath, and it is still an open question to identify realistic defects in oxide materials. Encouragingly, recent first-principles simulations discovered a Bi complex in CaO, with optical transitions in the telecommunication regime and a coherence time of 4.7 s at clock transitions \cite{davidsson2024discovery}.  

In this work, we identify and characterize a promising spin defect in MgO, which  has a long and rich history in the microelectronics industry, particularly in spintronics. Magnesium oxide plays a key role in enabling giant tunnel magnetoresistance in magnetic tunnel junctions used in random access memories \cite{butler2001spin, mathon2001theory, yuasa2004giant, parkin2004giant, yuasa2007giant}. In addition, MgO tunnel barriers enhance spin injection efficiency from the ferromagentic source to the semiconducting channel in spin field effect transistors \cite{jiang2005highly}. Although controversial, it has been suggested that MgO might exhibit d$^0$-ferromagnetism \cite{kapilashrami2010experimental, li2013origin, singh2017d}, which could enable the use of the material as a dilute magnetic semiconductor spin filter in spin transistors. We note that the high dielectric constant of MgO, its low leakage current, and thermal stability make it an ideal material for use as a gate dielectric in thin film transistors \cite{jiang2016solution}. Furthermore, MgO is CMOS-compatible and can be grown in the form of high quality thin films via a variety of methods \cite{plociennik2016optical, valanarasu2014microstructural}, and it is frequently used as a template layer for ferroelectrics \cite{basit1998growth, murphy2004electronic} and superconductors \cite{groves1999ion}. Thus, identifying spin defects in MgO could open avenues to combine spintronics, ferroelectrics, and quantum information science for the realization of multifunctional, hybrid classical-quantum applications.

Utilizing a high-throughput framework and advanced techniques, including hybrid time-dependent density functional theory (TDDFT) and quantum defect embedding theory (QDET), we identify and characterize a promising spin defect in MgO: an NV$^-$-center. We find that the ground and first excited state of this defect are triplet, with intermediate singlet shelving states; in addition the NV$^-$ in MgO exhibits a large zero-field splitting (ZFS), and a broad emission side-band in the visible regime. All of these properties together point towards its potential use as a quantum sensor or a spin qubit for quantum technologies. 
In our study we also report emission spectra, and computed ZFS and hyperfine parameters that may enable the experimental identification of the predicted defect; additionally we  analyze the reasons behind the strong vibronic coupling of the electronic states of the NV$^-$ defect with the host MgO lattice, thus providing insights into the engineering of spin defects in oxides. 

\section{Results}

\subsection{High-throughput screening of spin defects in MgO}

\begin{figure}
    \centering
    \includegraphics[width=\columnwidth]{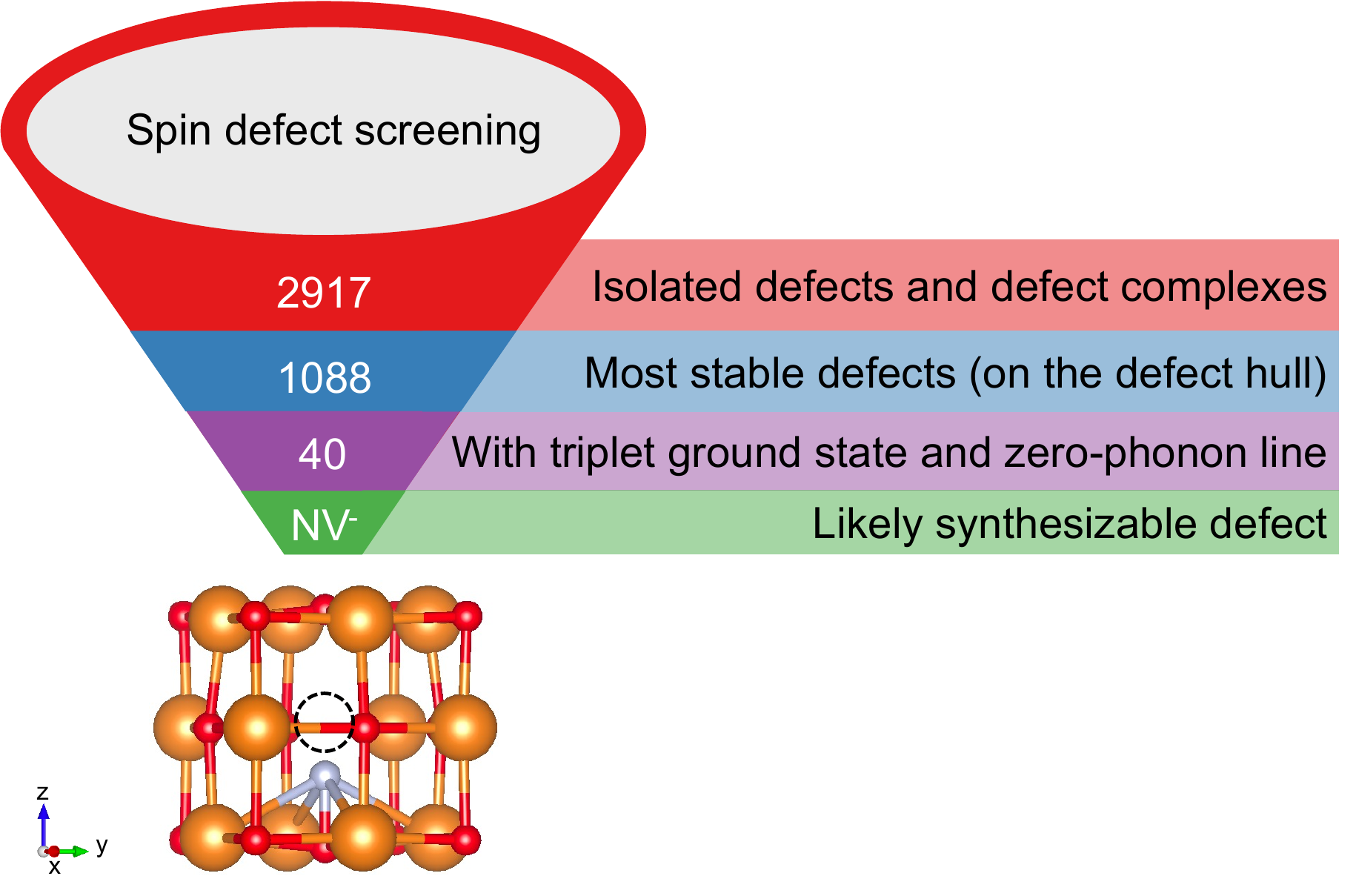}
    \caption{\justifying \textbf{High-throughput screening of spin defects in MgO.} Schematic of the workflow, which includes creation of isolated defects and complexes including $s$- and $p$-dopants and intrinsic defects, which are screened based on stability, presence of a spin triplet ground state and zero-phonon line, and synthesizability (see text), leading to the identification of an NV$^-$ center in MgO (shown in inset).}
    \label{fig:ht}
\end{figure}

We used a high-throughput technique to screen through thousands of isolated defects and defect complexes and identify favorable spin defect candidates in MgO. The process followed in our study is summarized in Fig. \ref{fig:ht}.
The screening was performed with the  ADAQ (Automatic Defect Analysis and Qualification) \cite{ADAQ,adaq_info} framework, which in turn uses the high-throughput toolkit (\textit{httk}) \cite{armientoDatabaseDrivenHighThroughputCalculations2020}.
This framework has been successfully applied to several host materials, including diamond \cite{davidsson2024diamond}, SiC \cite{modvac}, and CaO \cite{davidsson2024discovery}.

In our search, we considered isolated defects and defect complexes (the latter comprising two neighboring individual defects), including native vacancies and interstitials and their complexes, and substitutional and interstitial defects and defect complexes with $s$- and $p$-elements of the periodic table. We limited our investigation to $s$- and $p$-dopants as we use DFT at the PBE level to screen defects, and we expect that $d$- and $f$-dopants would require the use of DFT+U or hybrid DFT methods, which will be the topic of future investigations, given their higher computational cost. 
For simplicity, we excluded interstitial-interstitial complexes from the screening.
We also restricted our search to complexes with first or second nearest-neighbor defect constituents (i.e. the maximum distance between the two constituent defects in a complex is 3.5 $\mathrm{\AA}$), given the cell sizes that are reasonably affordable in our high-throughput procedure, and we included a single type of extrinsic dopant (i.e. we do not consider defect complexes of the type $\mathrm{X_{Mg}Y_{Mg}}$, $\mathrm{X_{Mg}Y_{O}}$, $\mathrm{X_{Mg}Int_{Y}}$, $\mathrm{X_{O}Int_{Y}}$, where X and Y are different elements substituting the Mg or O site or existing as an interstitial (Int)).
Our initial screening yielded 2917 defects, 
of which 1088 were determined to be on the defect hull--i.e. to have the lowest formation energy per stoichiometry and per Fermi energy \cite{modvac}. 
Out of those, we considered the defects  with a stable spin triplet ground state (which can thus serve as a two-level system) and which possess a zero-phonon line (ZPL), arriving at the list of the 40 defects tabulated in Supplementary Tables 1 and 2.

We found that 24 of the 40 defects are complexes comprised of second nearest-neighbor defects, which are excluded from further analysis as their experimental synthesis/positioning will likely be challenging. The remaining isolated defects and first nearest-neighbor complexes are shown in 
Table \ref{tab:spin-1}. We find that 
$\mathrm{Ga_O}$ is the only isolated defect with a stable spin triplet ground state;
however, we did not consider it since $\mathrm{Ga_{Mg}}$ has a much lower formation energy and thus it is more likely to be formed when $\mathrm{Ga}$ is implanted in MgO.
Several complexes listed in Table \ref{tab:spin-1} are of the type $\mathrm{X_OVac_{Mg}}$, where the dopant X (X = B, I, Sb, Bi) lies between the magnesium and oxygen vacancy sites.
This class of defects was first identified in CaO, where the Bi dopant has been proposed to be suitable for quantum applications due to the presence of a clock transition \cite{davidsson2024discovery}. These defects were excluded from the present study, as we expect them to have similar optical and spin properties as those in CaO, including, unfortunately, a low Debye-Waller factor. Note that in MgO, the $\mathrm{B_OVac_{Mg}}$ defect is stable, unlike in CaO,  but it has a higher formation energy (by ~7 eV) than those discussed below. 
Further, several complexes of the type $\mathrm{X_{Mg}X_O}$, where X = B, Al, Ga, In, Tl are stable, with a detectable ZPL; however they are likely difficult to synthesize due to the slow diffusion of substitutional impurities \cite{davidsson2024diamond} and high formation energies (due to cations substituting the oxygen site), and hence they were not further analyzed.
Interestingly, our search identified additional five defects with a spin triplet ground state and possessing a ZPL; but they all consist of anions substituting the magnesium site and/or cations substituting the oxygen site, and thus they will likely have large formation energies and be challenging to synthesize (see Supplementary Note 1).

Finally, we are left with the defect $\mathrm{Int_NVac_{Mg}}$, comprising a nitrogen interstitial next to a magnesium vacancy; it is similar to  the NV$^-$ in diamond, which however is formed by a nitrogen substitutional, instead of interstitial,  next to a carbon vacancy. 
The synthesis of $\mathrm{Int_NVac_{Mg}}$ is expected to be straight-forward, since interstitials generally have low diffusion barriers in MgO \cite{el2018effect, brudevoll1996interstitial}.  
Moreover, nitrogen doping of MgO via, e.g. molecular beam epitaxy and implantation has been reported in the literature \cite{yang2010ferromagnetism,liu2011optical,li2013room,wang2021room,grob2012catalytic,li2013annealing}.
Thus, we choose to investigate the $\mathrm{Int_NVac_{Mg}}$ defect (referred to as NV$^-$ hereafter) for the remainder of our study.

We note that several previous studies  have investigated nitrogen dopants in MgO using first-principles methods \cite{zhang2011ab, slipukhina2011ferromagnetic, mavropoulos2009ferromagnetism, pardo2008magnetism, pesci2010nature, polfus2012nitrogen, bannikov2007novel}, addressing the possibility of d$^0$-magnetism in MgO  and  its application as a dilute magnetic semiconductor for spintronic devices. These studies explored the effect of nitrogen substituting the oxygen site \cite{zhang2011ab, mavropoulos2009ferromagnetism, pardo2008magnetism, pesci2010nature, polfus2012nitrogen, bannikov2007novel}, nitrogen interstitials \cite{pesci2010nature}, and nitrogen dimers \cite{slipukhina2011ferromagnetic}. However, they did not investigate the NV$^-$ complex, and do not report any excited state properties relevant to quantum applications.  

\begin{table}[h!]
\caption{\justifying Isolated defects and first nearest-neighbors complexes on the defect hull, with spin triplet ground state and zero-phonon line (ZPL), as obtained from the screening process of Fig.\ref{fig:ht}, using the PBE functional.}
\begin{tabular} {cc|r|r}
\hline
\hline
Defect & Defect & Charge & ZPL \\
Type & & & [eV] \\
\hline
\multirow{4}{*}{$\mathrm{X_OVac_{Mg}}$}
& $\mathrm{B_OVac_{Mg}}$ & -1 &  0.5    \\
& $\mathrm{I_OVac_{Mg}}$  &  1   &  0.74  \\
& $\mathrm{Sb_OVac_{Mg}}$ &  -1 &  3.1   \\
& $\mathrm{Bi_OVac_{Mg}}$ &  -1 &  0.7  \\

\hline
\multirow{5}{*}{\shortstack{$\mathrm{X_{Mg}X_O}$}}
 & $\mathrm{B_{Mg}B_O}$ &  0  &  1.39  \\
 & $\mathrm{Al_{Mg}Al_O}$ &  0  &  0.68    \\
 & $\mathrm{Ga_{Mg}Ga_O}$ &  0    &  1.29     \\
 & $\mathrm{In_{Mg}In_O}$  &  0   &  0.98    \\
 & $\mathrm{Tl_{Mg}Tl_O}$   &  0   &  1.81   \\

\hline
\multirow{1}{*}{$\mathrm{X_O}$}
 & $\mathrm{Ga_O}$ &  -1   &  4.19  \\

\hline
\multirow{1}{*}{$\mathrm{Int_XVac_{Mg}}$}
& $\mathrm{Int_NVac_{Mg}}$  &  -1   &  2.19  \\

 \hline
 \hline
\end{tabular}
\label{tab:spin-1}
\end{table}

\subsection{Ground state properties using hybrid DFT}

We investigated the ground state properties of the NV$^-$ center in MgO using density functional theory (DFT) with the dielectric-dependent hybrid (DDH) exchange-correlation functional \cite{skone2014self}. The inverse of the experimental macroscopic dielectric constant $\epsilon_\infty= 2.96$ was used as the fraction of exact exchange, i.e. $\alpha$ = 0.34 \cite{lide2004crc, skone2014self}. The DDH functional is able to recover the experimental band gap at 0 K accurately (after accounting for zero-point motion renormalization effects), unlike semi-local functionals like PBE, which underestimate the bandgap (see Supplementary Table 3 for a comparison between the band gaps obtained using PBE, DDH, and experiment). 

\begin{figure*}
    \centering
    \includegraphics[scale=0.5]{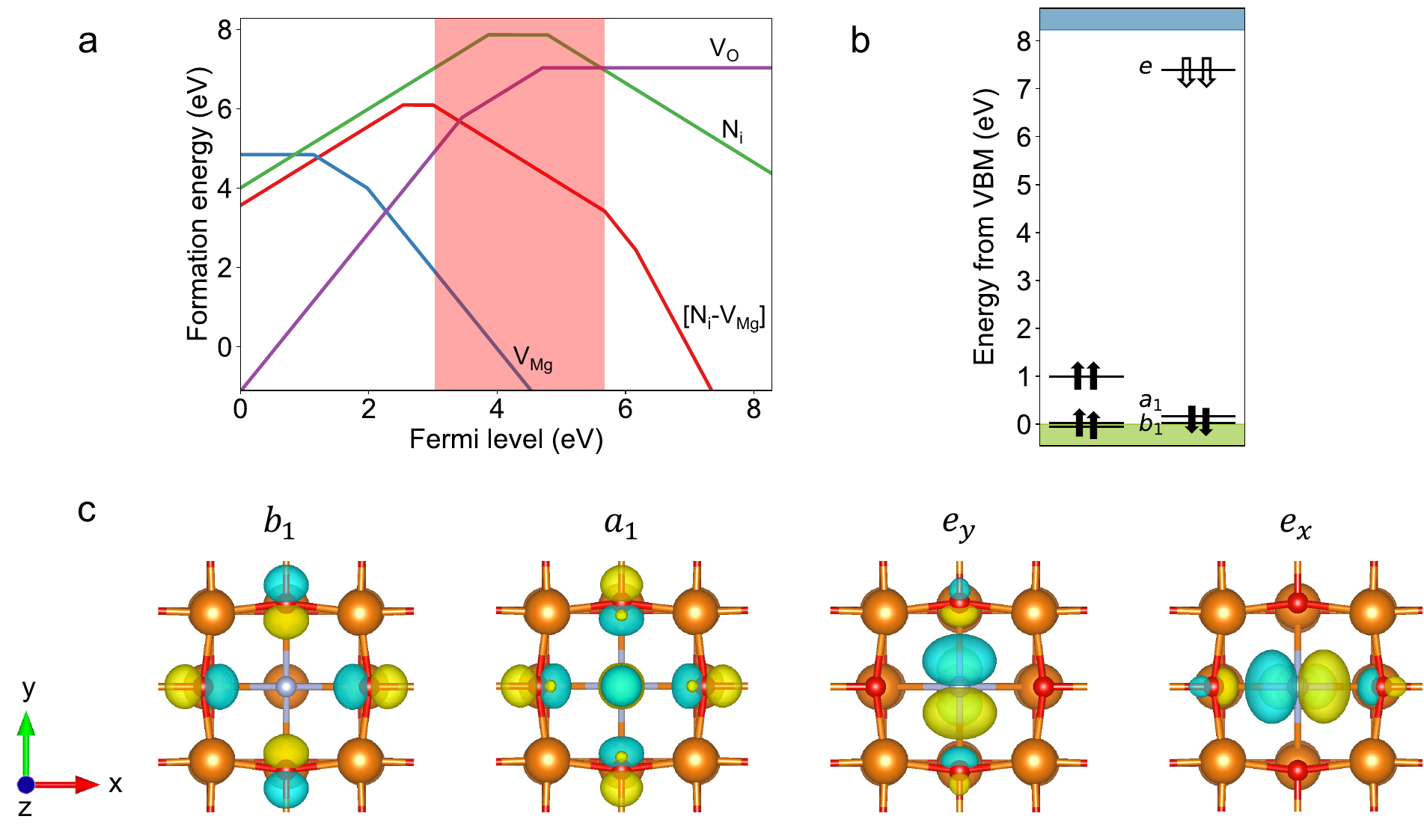}
    \caption{\justifying \textbf{Ground state properties of the NV$^-$ center in MgO computed with  hybrid DFT (see text) a.} Formation energies of different point defects in MgO under oxygen- and nitrogen-rich conditions as a function of the Fermi level: $\mathrm{V_O}$, $\mathrm{N_i}$, $\mathrm{V_{Mg}}$ and $\mathrm{[N_i-V_{Mg}]}$ denote an oxygen vacancy, nitrogen interstitial, magnesium vacancy and a nitrogen interstitial-magnesium vacancy complex, respectively. The shaded area shows the stability region for the NV$^-$ defect. \textbf{b.} Defect level diagram of the NV$^-$ defect with the single-particle defect levels labeled according to the irreducible representation of the $C_{4v}$ point group. Solid (empty) arrows denote occupied (unoccupied) defect levels. In the minority spin channel, the $b_1$, $a_1$, and $e$ orbitals are 0.03 eV, 0.17 eV, and 7.39 eV above the VBM, respectively. \textbf{c.} Iso-surfaces (level: 0.003 e/\AA$^3$) of the square moduli of the Kohn-Sham orbitals in the spin down channel. Orange, red and gray spheres represent magnesium, oxygen and nitrogen atoms, respectively. Viewing direction is along the $C_{4}$ axis.}
    \label{fig:gs-prop}
\end{figure*}

As shown in the inset of Fig. \ref{fig:ht}, the NV$^-$ defect in MgO has $C_{4v}$ symmetry in the ground state, with a nitrogen atom present at the bond-center interstitial site, bonded to four neighboring magnesium atoms (d$_{\mathrm{N}-\mathrm{Mg}}$ = 2.49 \AA) and an oxygen atom (d$_{\mathrm{N}-\mathrm{O}}$ = 1.33 \AA), with a nearest neighbor magnesium vacancy and an additional electron. The fourfold rotation axis ($C_4$) is along the N-O bond in the $<$001$>$ direction. We plot the defect formation energies of relevant defects under O$_2$-rich, N$_2$-rich conditions as a function of the Fermi level in Fig. \ref{fig:gs-prop}a. We see that the NV$^-$ is stable within the bandgap, and therefore, may be experimentally realizable. Further, we expect that NV$^-$ will be compensated by positively charged oxygen vacancies. Note that the Fermi energy may be pinned at 2.3 eV above the valence band maximum (VBM) due to compensating native defects (magnesium and oxygen vacancies) under high temperature equilibrium growth conditions. One strategy to move the equilibrium Fermi level into the stability region of the NV$^-$ defect could be to $n$-dope MgO during growth, for example, with Al$^{3+}$ \cite{hong2009effects} or  Fe$^{3+}$ \cite{perez1983iron}. This would have to be done carefully to avoid the introduction of additional optically active defects and a high concentration of spinful nuclei into the sample. However,  sample preparation techniques could be designed to circumvent the requirement of $n$-type doping. Specifically, sample preparation typically involves high temperature growth in a nitrogen-rich environment, followed by a cooling down process where the nitrogen flux is turned off, thereby fixing the concentration of nitrogen in the sample. Under these low temperature conditions and high nitrogen concentrations, where the contribution from configurational entropy is low, binding energy effects will be predominant \cite{van2004first}
; hence they will promote the formation of the NV$^-$ complex, especially given its high binding energy ($E_b$ = 2.86 eV, calculated as \(E_b =  E_f(V_{Mg}^{2-}) + E_f(N_i^{1+}) - E_f(NV^-)\); $V_{Mg}^{2-}$ and $N_i^{1+}$ denote the doubly-negatively charge magnesium vacancy and the singly-positively charged nitrogen interstitial). In general, the defects formed will strongly depend on the sample preparation techniques (for example, nitrogen incorporation during growth vs. implantation). Fig. \ref{fig:gs-prop}a is indicative of the fact that the NV$^-$ defect can be formed in MgO by optimizing sample preparation conditions, as it is stable within the bandgap. Additionally, we checked for the possibility of the nitrogen interstitial recombining with the magnesium vacancy to form a substitutional nitrogen defect on the magnesium site. We found that the NV$^-$ defect is the preferred configuration; this result can be understood from a molecular orbital analysis, presented in Supplementary Note 3. Finally, we also computed the barrier for the nitrogen interstitial to migrate to other interstitial sites around the vacancy (see Supplementary Note 4). The computed barrier is high ($\approx1.37$ eV), thus, once formed, the NV$^-$ defect is expected to be stable against annealing. 

The defect level diagram of the NV$^-$ center obtained using the DDH functional is shown in Fig. \ref{fig:gs-prop}b. We see that the ground state is a spin triplet, with four isolated defect states present within the bandgap of MgO: one $b_1$, one $a_1$, and two degenerate $e$ orbitals. Fig. \ref{fig:gs-prop}c shows the iso-surfaces of the square moduli of the Kohn-Sham orbitals of the four defect states in the spin down channel as viewed along the $C_4$ axis. We find that the $b_1$ and $a_1$ defect states near the valence band maximum (VBM) consist of O 2$p$ orbitals arising from the presence of the magnesium vacancy, and the degenerate $e$ defect states consist of $\pi$-orbitals ($\pi_{px}$, $\pi_{py}$) from the N-O bond. In fact, the additional electron in this complex localizes on the N-O bond in one of the $\pi$-orbitals. The separation of the occupied and unoccupied defect states from each other as well as from the valence and conduction bands of MgO indicate that optical transitions to prepare and measure the qubit state are possible.    

To aid in the experimental identification of the NV$^-$, for example via electron paramagnetic resonance (EPR) or magnetic circular dichroism experiments, we calculated the zero-field splitting (ZFS) tensor and the hyperfine parameters. A sufficiently large ZFS is required to isolate the $m_s$ = $\pm$1 and $m_s$ = 0 sublevels of the ground state triplet at zero magnetic field, allowing for the controllable preparation of the spin qubit. As expected from the $C_{4v}$ symmetry of this defect, the transverse $E$ value is 0. Our computed axial $D$ component of the ZFS tensor is 38.48 (46.38) GHz at the PBE (DDH) level. This value represents the energetic separation between the $m_s$ = $\pm$1 and $m_s$ = 0 sublevels, indicating that the defect can be initialized conveniently in the $m_s$ = 0 sublevel. Furthermore, the computed value of $D$ is favorable to prevent excitations from the $m_s$ = 0 to $m_s$ = $\pm$1 sublevel, thus protecting the qubit from decoherence, while still being measurable via traditional microwave EPR experiments (instead of requiring a high-frequency EPR setup, for example). Note that this value of ZFS is much larger than that reported for defect qubits in diamond, SiC, and AlN (where many spin defects  have ZFS smaller than 5 GHz \cite{seo2017designing}), but it is of the same order of magnitude as that of the $\mathrm{Ti_{VV}}$ center in 2D h-BN (19.4 GHz) \cite{smart2021intersystem} and SiV in diamond (46 GHz) \cite{meesala2018strain}, and lower than that reported for other qubit systems, e.g. single molecule magnets (for which ZFS can reach up to 13000 GHz) \cite{schulte2018effects}.  

EPR measurements also yield hyperfine parameters, which describe the coupling between the electron spins and nuclear spins. The NV$^-$ defect in MgO has the following nuclear spins: $^{14}$N ($I$ = 1, 99.63$\%$), $^{25}$Mg ($I$ = 2.5, 10$\%$), and $^{17}$O ($I$ = 2.5, 0.038$\%$), and the principal values of its hyperfine tensors are given in Table \ref{tab:hfi}. Pesci et. al \cite{pesci2010nature} investigated nitrogen interstitials in MgO and  found that  electrons localize on the $\pi$-orbitals around the N-O bond, consistent with our findings. Interestingly, their reported hyperfine parameters for substitutional nitrogen on the oxygen site and interstitial nitrogen are different from our calculated values for the NV$^-$ defect (see Supplementary Table 4 for a comparison), indicating that the latter might be experimentally distinguishable from other nitrogen defects in the MgO lattice, using EPR.

Overall, the ground state properties of the NV$^-$ in MgO indicate that this defect is stable over a large energy range within the band gap of the host and thus, should be experimentally realizable. It has a spin triplet ground state with isolated occupied and unoccupied levels, pointing to the possibility of optical transitions to prepare and measure a qubit state. The computed ZFS tensor shows that a sufficiently large separation between spin sublevels is present, and that initialization in the $m_s$ = 0 sublevel should be realizable experimentally. Further, the computed hyperfine parameters indicate that the NV$^-$ center is expected to be distinguishable from other nitrogen defects in MgO.  

\begin{table}[h]
    \caption{\justifying Computed principal values of the hyperfine tensors (A) at the PBE level of theory (in MHz) for the NV$^-$ defect in MgO.}
    \centering
    \begin{tabular}{cccc}
    \hline
    \hline
        Atom & A$_{xx}$ & A$_{yy}$ & A$_{zz}$ \\
        \hline
        N atom & 43.58 & 43.59 & -87.17 \\
        O atom bonded to N & -27.86 & -27.89 & 55.76 \\
        Mg atoms nearest-neighbor to N & 0.27 & 0.47 & -0.74 \\
        \multirow{2}{*}[4pt]{O atoms nearest-neighbor} & \multirow{2}{*}{11.07} & \multirow{2}{*}{12.30} & \multirow{2}{*}{-23.37} \\
        to magnesium vacancy& & & \\
        \hline
        \hline
    \end{tabular}
    \label{tab:hfi}
\end{table}

\subsection{Excited state properties using hybrid TDDFT}

\begin{figure*}
    \centering
    \includegraphics[scale=0.5]{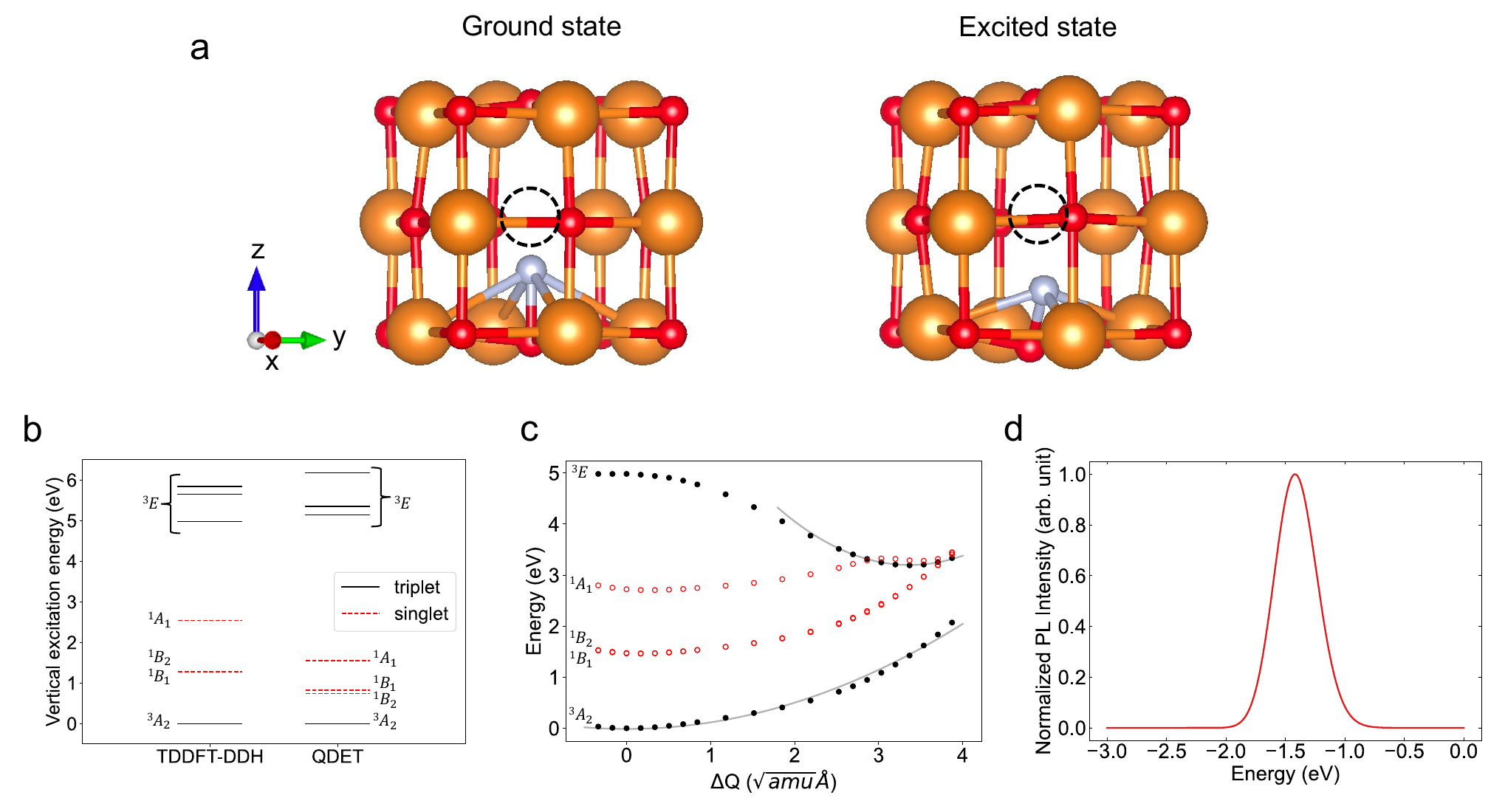}
    \caption{\justifying \textbf{Excited state properties of the NV$^-$ center in MgO a.} Optimized structures of the ground state $^3A_2$ and first excited $^3E$ state obtained using DFT-DDH and TDDFT-DDH, respectively. Black dotted circle denotes the magnesium vacancy. \textbf{b.} Many-body level structure obtained using TDDFT-DDH and quantum defect embedding theory (QDET). \textbf{c.} One-dimensional configurational coordinate diagram constructed between the optimized structures of the $^3A_2$ ground state and first $^3E$ excited state. The points denote energies obtained at each ground state (excited state) configuration using single-point DDH (TDDFT-DDH) calculations. Solid lines show a quadratic fit. \textbf{d.} Calculated emission (photo-luminescence, PL) spectrum at 300 K aligned with respect to the zero-phonon line.}
    \label{fig:es-prop}
\end{figure*}

We now turn to investigate the excited state properties of the NV$^-$ center using time-dependent density functional theory (TDDFT) with the DDH functional, as implemented in the WEST code \cite{jin2023excited}. We used spin-conserving and spin-flip TDDFT-DDH to investigate triplet and singlet excited states, respectively. The relaxed structure of the triplet ground state and first excited triplet state (obtained using DDH and TDDFT-DDH, respectively) are shown in Fig. \ref{fig:es-prop}a.  

Fig. \ref{fig:es-prop}b plots the vertical excitation energies of multiple low-lying singlet and triplet states, as obtained from TDDFT-DDH. The first excited state triplet $^3E$ is about 5 eV above the triplet ground state ($^3A_2$), with additional excited state triplets ($^3E$) close in energy. Additionally, we find three singlet states ($^1A_1$, $^1B_1$, $^1B_2$) within the energy range of the first triplet excitation, indicating that an optical initialization cycle, similar to that of the NV$^-$ center in diamond, may be possible. 

We also characterized the electronic excitations using quantum defect embedding theory (QDET) \cite{ma2021quantum, sheng2022green, vorwerk2022quantum} and compared the results with those of TDDFT-DDH. In QDET, we define an active space that consists of single-particle orbitals involved in the excitations of interest and a many-body effective Hamiltonian acting on this space, which includes the influence of the environment surrounding the defect. The electronic structure of the environment is described at the $G_0W_0$ level of theory, enabling an exact correction for the double counting terms. For the NV$^-$, we chose an active space consisting of the four localized defect states within the band gap ($b_1$, $a_1$, $e_x$, $e_y$), as well as of defect states close to the valence band maximum. The eigenstates of the effective Hamiltonian are determined exactly using the full configuration interaction method, yielding the many-body level diagram shown in Fig. \ref{fig:es-prop}b. We see that the trends in excited state energies using TDDFT-DDH and QDET are the same. The minor differences in absolute energy values between the two methods could arise due to the absence of double excitations in TDDFT and/or a limited active space in QDET. Work is ongoing to understand the performance of QDET on a broader class of defects and hosts, which will help clarify the reasons behind the difference between the TDDFT and QDET results; detailed comparisons will be reported elsewhere. For the problem at hand here, the important point is that the order of multi-reference states obtained with TDDFT and QDET is the same.

Having determined the vertical excitation energies, we proceeded to relax the structure in the first excited triplet ($^3E$) state using TDDFT-DDH with analytical forces \cite{jin2023excited} (see Fig. \ref{fig:es-prop}a). A decomposition of the many-body state shows that the $a_1$ $\rightarrow$ $e$ excitations are dominant. We note that in the relaxed geometry of the excited state, the symmetry is reduced to $C_1$, with d$\mathrm{_{N-Mg}}$ = 1.98 - 2.86 \AA  and d$\mathrm{_{N-O}}$ = 1.37 \AA. We find four equivalent minima, with the N-O bond tilted along the $+x$, $-x$, $+y$, $-y$ directions, consistent with the $C_{4v}$ symmetry of the ground state. Table \ref{tab:op-prop} lists the absorption, zero phonon line (ZPL), and emission energies of the NV$^-$ in MgO (calculated under the Franck-Condon approximation) and compares it with that of the NV$^-$ center in diamond and the VV$^0$ in silicon carbide. In MgO, the absorption and ZPL of the NV$^-$ center are in the ultra-violet (UV) range whereas the emission is in the visible range, highlighting a large Stokes shift. 

\begin{table}
    \caption{\justifying Optical properties of the NV$^-$ defect in MgO obtained using TDDFT-DDH (see text). The values for diamond and SiC are taken from the literature and are obtained using the HSE functional. The two values for SiC are for the $hh$ axial $\mathrm{VV^0}$ and $hk$ basal $\mathrm{VV^0}$ respectively. ZPL: zero-phonon line, $\Delta R$: total displacement between the excited and ground state structures, $\Delta Q$: mass-weighted displacement between the excited and ground state structures.}
    \centering
    \begin{tabular}{cccc}
    \hline
    \hline
    Property & NV$^-$ MgO & NV$^-$ diamond & VV$^0$ 4H-SiC \\
    \hline
    Absorption (eV) & 4.98 & 2.21\cite{gali2009theory} & 1.18, 1.27\cite{gordon2015defects} \\
    ZPL (eV) & 3.19 & 1.96\cite{gali2009theory} & 1.13, 1.21\cite{gordon2015defects} \\
    Emission (eV) & 1.76 & 1.74 \cite{gali2009theory} & 1.09, 1.18 \cite{gordon2015defects} \\
    $\Delta R$ (\AA) & 0.84 & 0.18 \cite{alkauskas2014first} & 0.16, 0.18 \cite{ADAQ}  \\
    $\Delta Q$ (amu$^{0.5}$\AA) & 3.37 & 0.62\cite{gali2009theory} & 0.60, 0.74 \cite{ADAQ}  \\
    \hline
    \hline
    \end{tabular}
    \label{tab:op-prop}
\end{table}

A large Stokes shift implies that a large number of phonons are involved in the emission process, which could lower the intensity of the ZPL relative to the rest of the emission spectrum, thus yielding an unfavorable Debye-Waller factor (DWF). The average number of phonons emitted during an electronic transition is called the Huang-Rhys factor (HRF, denoted as $S$) and is given by \cite{stoneham2001theory}

\begin{equation}\label{hrf}
        S = \frac{\frac{1}{2}\Delta{Q}^2\omega_{eff}^2}{\hbar\omega_{eff}}
\end{equation}

where $\omega_{eff}$ is an effective phonon frequency and $\Delta Q$ a mass-weighted displacement. The latter is calculated as 

\begin{equation}
\Delta Q = (\sum_{\alpha=1}^N \sum_{i=x,y,z} M_\alpha \Delta R_{\alpha i}^2)^{1/2}
\end{equation}

where $M_\alpha$ = mass of atom $\alpha$, $\Delta$$R_{\alpha i}$ = displacement of atom $\alpha$ between the excited state and ground state equilibrium structures in the $i^{th}$ direction). Not surprisingly, given the large Stokes shift, we find that the $\Delta Q$ of the NV$^-$ is large (see Table \ref{tab:op-prop}),  indicating a strong electron-phonon coupling \cite{stoneham2001theory, Bersuker_2006}.  

We calculated $\omega_{eff}$, and therefore, the HRF $S$, by constructing a one-dimensional configuration coordinate diagram (CCD) along the direction connecting the relaxed $^3A_2$ ground and $^3E$ excited state structures \cite{jin2021photoluminescence, alkauskas2014first} (geometries optimized using DDH and TDDFT-DDH, respectively), as shown in Fig. \ref{fig:es-prop}c. By fitting a quadratic function around the points near the respective minima, we obtain the effective phonon frequencies of the ground and excited states to be $\hbar\omega_g$ = 32.79 meV and $\hbar\omega_e$ = 61.77 meV respectively. Using $\hbar\omega_g$ and $\Delta Q$, we obtain the HRF for emission as $S$ = 44.49, which is again indicative of a strong electron-phonon coupling \cite{stoneham2001theory}, as expected. Such a large HRF leads to a negligible Debye-Waller factor (DWF $\simeq e^{-S}$ $\approx$ 10$^{-20}$), as a majority of the photons contribute to the emission side band instead of the ZPL. Thus, the ZPL will likely be undetectable, rendering the defect unsuitable as a single photon emitter (for example, for quantum communication applications). However, the defect's triplet ground state, large ZFS, and many-body level structure may make it suitable for quantum sensing, computing, or transduction, by enabling optical initialization and spin-selective photo-luminescence. In fact, the possibility of optical initialization and readout is further highlighted from the CCD of the singlet excited states plotted in Fig. \ref{fig:es-prop}c. We see that at the $^3E$ minimum, $^1A_1$ is nearly degenerate with $^3E$, and $^1B_1$ and $^1B_2$ are lower in energy than the $^3E$ state. Thus, despite the large $\Delta Q$ in this system, the singlet states' potential energy surfaces do not cross that of the triplet $^3E$ excited state (or are nearly degenerate with the $^3E$ minimum, as in the case for $^1A_1$), indicating that spin-selective photoluminescence via inter-system crossing may be possible. While the rates for these transitions would need to be calculated, this is an important prerequisite for the possibility of optical initialization and readout. Finally, nanostructuring or the additional presence of extended defects can be used to lower the HRF and therefore increase the intensity of the ZPL, as explored in SiC \cite{lee2021strong}.   

Using $\hbar\omega_g$, $\hbar\omega_e$, and $\Delta Q$, we show  the emission spectrum of the NV$^-$ in Fig. \ref{fig:es-prop}d, which is nearly a Gaussian with a peak at the Franck-Condon energy of 1.43 eV, and a line-width of 420 meV. The emission spectrum thus lies in the visible red-infrared region, making it easy to detect. Moreover, a broad line-width is advantageous as it would allow multiple frequency filters to be used (for example, to filter out emissions from an underlying substrate) while retaining a significant fraction of the emission spectrum to be employed for detection. The radiative lifetime is 23 $\mu$s, calculated as the inverse of the radiative rate \(\Gamma_R = \frac{n_r \mu_{\text{em}}^2 (E_{\text{em}})^3}{3 \pi \epsilon_0 c^3 \hbar^4}\), where $n_r$ is the refractive index of MgO (1.73), $\mu_{em}$ is the transition dipole moment for emission (0.035 $e$\AA) calculated using TDDFT-DDH, and $E_{em}$ is the emission energy calculated using TDDFT-DDH (1.76 eV). This value is three orders of magnitude larger than those computed for the NV$^-$ in diamond \cite{gali2019ab} and VV$^0$ in SiC \cite{bian2024theory}, but is similar to those computed for spin defects in 2D transition metal dichalcogenides \cite{lee2022spin}. While it would be preferable to have a shorter radiative lifetime, the non-radiative lifetime is likely longer, as the $^3E$ and $^3A_2$ potential energy surfaces cross only at large barrier energies, and non-radiative rates are typically low for defects with transition energies larger than 1.5 eV \cite{turiansky2024rational}. Moreover, we note than the predicted coherence time $T_2$ for an electron spin defect in MgO is 600 $\mu$s \cite{kanai2022generalized}, thus, the spin state could still be retained during optical initialization and readout. Nonetheless, cavity enhancement may be required due to other competing processes like inter-system crossing. Cavity enhancement of over 1000 has recently been reported for transitions in Er$^{3+}$-doped MgO \cite{horvath2023strong}. This result points to the encouraging possibility of cavity enhancement of the transitions in the NV$^-$ center in MgO as well. 

To summarize the computed excited state properties of the NV$^-$ in MgO, we find  the presence of triplet excited and singlet shelving states, pointing at the possibility of an optical initialization cycle and spin-selective photo-luminescence. Further, our results show that the absorption and ZPL energies are in the UV range, whereas the emission energy is in the visible (red) range. The visible emission energy, together with the computed broad emission side-band, is a convenient feature for experimental detection in applications such as quantum sensing and transduction. We also find a large $\Delta Q$ and  related Stokes shift, large HRF, and low DWF, indicating strong vibronic coupling. Identifying the reason behind this strong coupling is key to determining engineering strategies that can mitigate it, and to enhance our understanding of oxides as hosts for spin defects.  We explore this in further detail below. 

\subsection{Vibronic analysis}

\begin{figure}[h]
    \centering
    \includegraphics[width=\columnwidth]{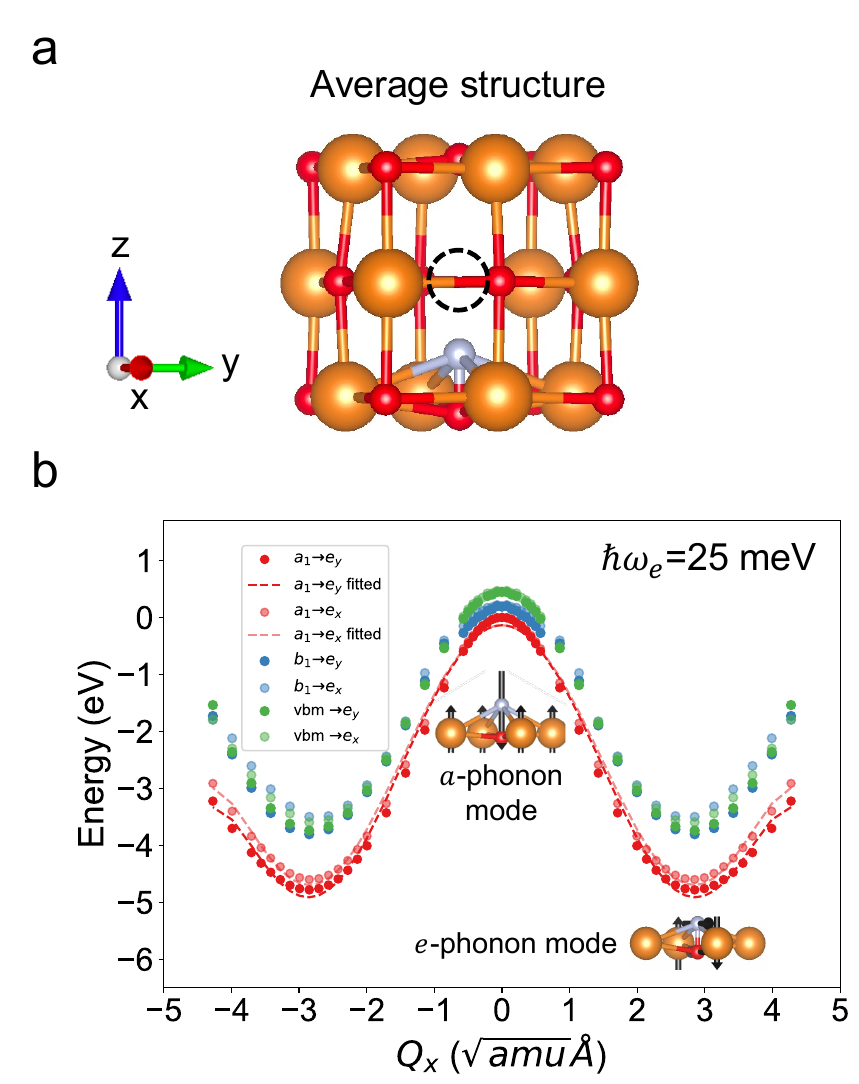}
    \caption{\justifying \textbf{Analysis of vibronic coupling a.} Average structure of the four equivalent excited state minima of the NV$^-$ center in MgO. \textbf{b.} One dimensional configurational coordinate diagram along $e_x$-phonon mode, showing the first three $^3E$ excited states. The points denote energies obtained at each configuration using single-point TDDFT-DDH calculations. The adiabatic potential energy surface of the first $^3E$-excited state is shown by the dashed lines. The predominant phonon modes involved in the average structure and the excited state structure are shown as insets. The displacement vectors of phonon modes are scaled up for clarity.}
    \label{fig:apes}
\end{figure}

According to the character table of the $C_{4v}$ point group (see Supplementary Table 5), the $E$ electronic states can couple linearly to the $a_1$, $a_2$, $b_1$, $b_2$ phonon modes, and to $e$ phonon modes via higher (even) order coupling. However, only the $b_1$, $b_2$, and $e$ phonon modes can reduce the symmetry of the defect. Given that the $C_{4v}$ symmetry is reduced to $C_1$ in the optimized geometry of the first excited $^3E$ state, $e$-phonon modes are expected to be predominantly coupled to the electronic states ($b_1$ and $b_2$ modes would have reduced the symmetry to $C_2$ and $C_{2v}$, respectively). To verify that indeed the $e$ phonon modes are responsible for the observed symmetry-breaking in the first excited $^3E$ state, we calculated the Mg, O, and N displacements between the average structure of the four equivalent minima (shown in Fig. \ref{fig:apes}a) and the optimized geometry of the excited state (shown in Fig. \ref{fig:es-prop}a). The average structure is symmetrical and  the displacement between the average geometry and that of the relaxed excited state geometry may be used to understand the predominant symmetry-breaking mode. We found that the symmetry-breaking displacements are predominantly due to the $e$-phonon mode (see Supplementary Note 6 and insets of Fig.\ref{fig:apes}b).

Thus, using the projections of the eigenvectors of the $e$-phonon modes as the collective variables (i.e. along $x$ and $y$ axes), and using the averaged structure as a reference geometry (since the symmetric $a_1$-phonon modes are not Jahn-Teller or pseudo-Jahn Teller active), we built a CCD of the first three excited states using TDDFT-DDH. The CCD along the $e_x$-phonon mode, identical to the one along the $e_y$ mode, is shown in Fig.\ref{fig:apes}b. Interestingly, we observe the presence of close-spaced excited states with negative curvature at $\Delta Q$ = 0, separated by only $\sim$0.2 eV, indicating that they likely interact strongly via the pseudo-Jahn Teller (pJT) effect. These excited states originate from excitations of the form $a_1$ $\rightarrow$ $e$, $b_1$ $\rightarrow$ $e$, and $VBM$ $\rightarrow$ $e$. The close-spaced excited states found here are $^3E$-states, similar to those seen in spin-conserving TDDFT-DDH and QDET calculations of the vertical excitation energies of the ground state structure, shown in Fig.\ref{fig:es-prop}b. 

To quantify the coupling between the various states, we fit an adiabatic potential energy surface (APES) to the first excited state CCD, which yields an effective phonon frequency of the $e$-phonon mode and the vibronic coupling constants. Specifically, we solve a $(E+E) \otimes e$ Hamiltonian, which accounts for Jahn-Teller (JT) coupling within each $^3E$ state and pseudo-Jahn Teller (pJT) coupling between two excited $^3E$ states. We used an adiabatic-to-diabatic transformation of the potential energy matrix and built an effective Hamiltonian that includes coupling constants for JT and pJT effects \cite{viel2004effects, eisfeld2005higher}. We then fit the eigenvalues of the Hamiltonian to the TDDFT-DDH energy values along the CCD of the first excited state to obtain the parameters of the energy matrix. The details of the derivation are given in Supplementary Note 7. The APES of the first excited $^3E$ state along the $e$-phonon mode is shown in Fig.\ref{fig:apes}b, from which we obtained the effective phonon frequency of the $e$-phonon mode to be 25 meV, and the vibronic coupling parameters to be 20-500 meV. We also found that the pJT effect is more pronounced than the JT effect: including the JT coupling parameters did not reduce the fitting error of the APES (see Supplementary Note 7 for more details). The pJT stabilization energy is large ($\sim$5 eV). These vibronic coupling parameters are far larger than those of the NV$^-$ in diamond (in which vibronic coupling is around 120 meV \cite{jin2022vibrationally}), and the effective phonon frequency is far lower than that of the NV$^-$ in diamond (effective phonon frequency of the latter is around 60 meV \cite{jin2022vibrationally}). The large pJT stabilization energy and low effective phonon frequencies satisfy the condition for strong vibronic coupling, given as \cite{Bersuker_2006}

\begin{equation}\label{strong_coupling}
    \frac{E_{PJT}}{\hbar\omega_{eff}} \gg 1
\end{equation}

Thus, from our vibronic coupling analysis, we find that the strong vibronic coupling of the first excited state to the $e$ phonon modes is due to a large pJT stabilization energy and low effective phonon frequencies. In turn, this strong coupling gives rise to a large $\Delta Q$ and then to a large Stokes shift and HRF, and to a low DWF. 

We end this section by briefly mentioning that the strong vibronic coupling could indicate a strong spin-phonon coupling, which might lead to a reduction in the spin relaxation time $T_1$. While a spin-phonon coupling analysis is beyond the scope of this study, we did calculate the phonon modes for the NV$^-$ defect in the ground state, as discussed in Supplementary Note 8. We find that the modes are largely localized near the NV$^-$ defect and have energies that generally exceed its ZFS. Thus, the spin relaxation time $T_1$ might be limited by slower two-phonon processes as opposed to fast one-phonon direct and Orbach relaxation mechanisms. Having predominantly two-phonon processes would be desirable for an optimal value of the defect's $T_1$.

\subsection{Engineering strategies to reduce vibronic coupling and increase DWF}

Having obtained insight into the reasons behind the strong vibronic coupling of the NV$^-$ center in MgO, we now discuss potential strategies to reduce such coupling, and therefore, the large $\Delta Q$, and potentially increase the DWF. 

As our first strategy, we explored the effect of strain on the optical properties of the NV$^-$. Strain has been reported to change the alignment of spins of the NV$^-$ in diamond \cite{karin2014alignment} and stabilize the spin triplet state in AlN \cite{seo2016design}.  We considered biaxial strain, since under thin film growth conditions, MgO will be epitaxially strained to match the lattice constants of the substrate, while relaxing its geometry along the perpendicular direction. Estimates using $\Delta$SCF-PBE (i.e. differences of total energies computed with different occupation numbers, see Supplementary Note 9) revealed that biaxial compressive strain in the $xy$-plane reduced the $\Delta Q$ while retaining the degeneracy of the $e$-levels. Hence, we investigated two different biaxial compressive strain levels: 1$\%$ and 4$\%$, and optimized the geometries of the NV$^-$ structure in its ground and first excited state using DFT-DDH and TDDFT-DDH, respectively. Our chosen strain values are experimentally representative, as MgO thin films can be epitaxially grown up to thicknesses of $\sim$1 nm with a biaxial compressive strain of 4$\%$ on substrates like Fe \cite{vassent1996study} and Ag \cite{valeri2002thickness}. Hereafter, `strain' refers to `biaxial compressive strain in the $xy$-plane'.

Notably, we find that as strain increases (0 \%, 1\%, 4\%), there is a reduction in $\Delta Q$ (3.37, 3.15, 2.78 amu$\mathrm{^{0.5}}$\AA), indicating a decrease in vibronic coupling.  

\begin{figure*}
    \centering
    \includegraphics[scale=0.5]{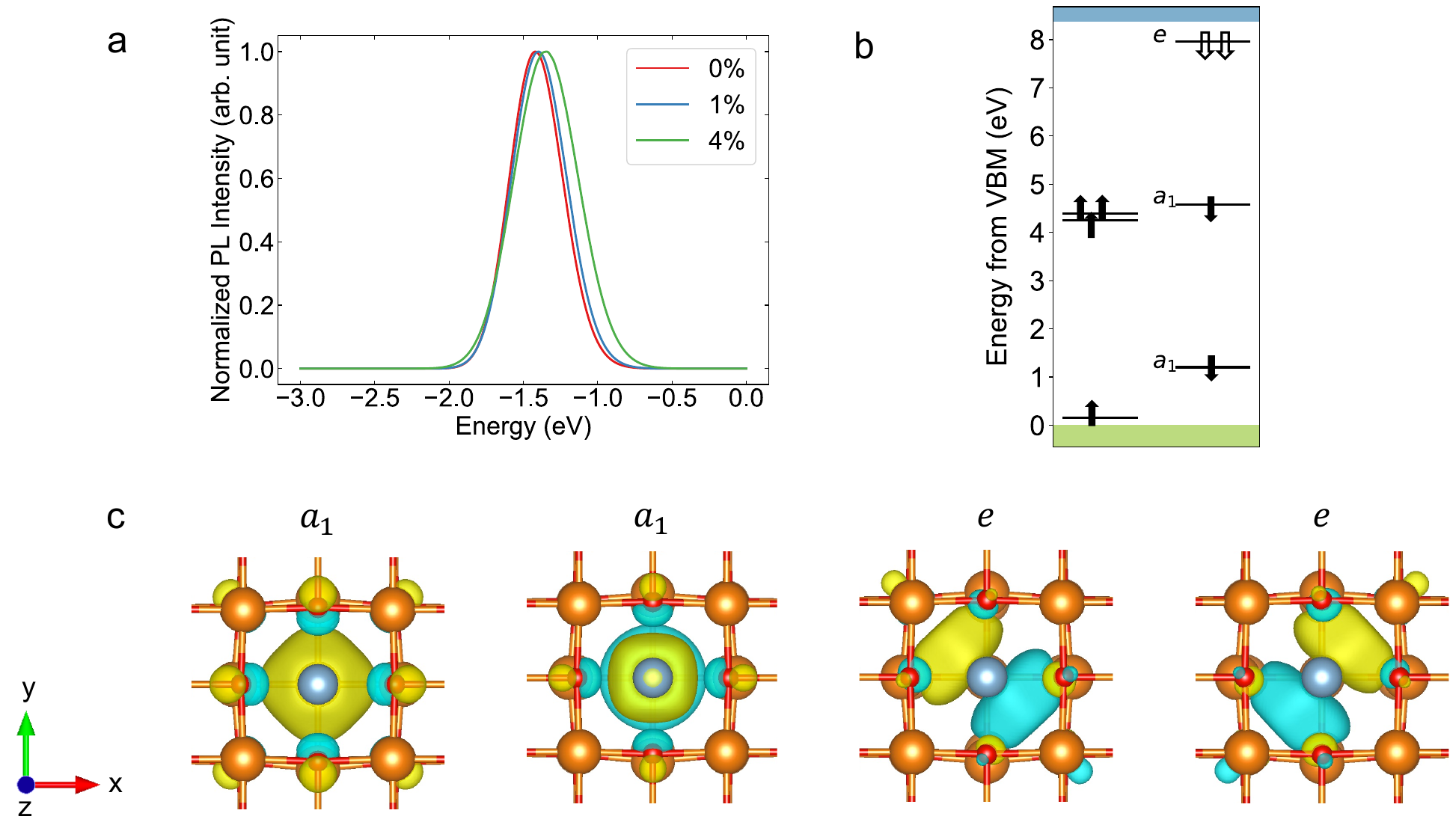}
    \caption{\justifying \textbf{Engineering strategies to reduce vibronic coupling a.} Calculated emission (photo-luminescence, PL) spectrum of the NV$^-$ defect in MgO at 300 K as a function of biaxial compressive strain, aligned with respect to the zero-phonon line. \textbf{b.} Defect level diagram obtained using the DDH functional for the $\mathrm{Al_{Mg}-Al_O}$ defect complex in MgO with the single-particle defect levels labeled according to the irreducible representation of the $C_{4v}$ point group. Solid (empty) arrows denote occupied (empty) defect levels. \textbf{c.} Iso-surfaces (level: 0.0007 e/\AA$^3$) of the square moduli of the Kohn-Sham orbitals in the spin down channel for the $\mathrm{Al_{Mg}-Al_O}$ defect complex.}
    \label{fig:engg-strat}
\end{figure*}

To check if this reduction in $\Delta Q$ leads to a reduced HRF and increased DWF, we followed the same strategy as described in Section C. We constructed a one-dimensional CCD along the direction connecting the optimized geometries in the ground and excited state using TDDFT-DDH and fit a quadratic function around the points in proximity of the respective minima. In this way, we obtained  effective phonon frequencies of the ground and excited states under the different strain conditions,  which we used to calculate the HRF. We find that strain decreases the HRF from 44.49 (unstrained case) to 41.36 (1\% strained case) and 36.23 (4\% strained case). This results in an increase in the DWF by 4 and 7 orders of magnitude respectively; however, unfortunately the ZPL would still be undetectable, as the DWF of the unstrained system is extremely low ($\approx$ 10$^{-20}$). 

The emission spectra at different strain conditions are shown in Fig.\ref{fig:engg-strat}a, calculated with the same method of Fig. \ref{fig:es-prop}d, i.e, by using the respective $\hbar\omega_g$, $\hbar\omega_e$, and $\Delta Q$. We once again see that the spectra are nearly Gaussian, with a mean around the respective Franck-Condon energies. The reduction in HRF is manifested as a slight red-shift in the mean of the emission spectrum.

Overall we find that strain can reduce the $\Delta Q$ and HRF, and therefore the vibronic coupling, thus increasing the DWF of the NV$^-$ defect in MgO by orders of magnitude; however, since the DWF of the unstrained system is negligible, the impact of strain on the DWF value is not sufficient to obtain a substantial increase. 

As our second strategy, we explored the role of the defect electronic structure on $\Delta Q$ (and therefore, vibronic coupling and DWF), by investigating another promising defect found in the MgO ADAQ database: the $\mathrm{Al_{Mg}-Al_O}$ defect complex. It has the the lowest $\Delta Q$ among the 20 nearest-neighbor defect complexes with a spin triplet ground state and ZPL, the latter in the near-telecom regime. In contrast to the NV$^-$ center, which consists of an interstitial-vacancy complex, the $\mathrm{Al_{Mg}-Al_O}$ complex consists of two substitutional impurities, and its analysis gives us insight into a different class of defects. Note however that unfortunately the $\mathrm{Al_{Mg}-Al_O}$ complex has a very high formation energy (around 11 eV); thus, its experimental realization might be challenging. 

Fig. \ref{fig:engg-strat}b shows the defect level diagram and Fig. \ref{fig:engg-strat}c shows the iso-surfaces of the square moduli of the Kohn-Sham orbitals of the four defect states in the minority spin channel. We note two key differences in the electronic structure of this defect, compared to the NV$^-$ center.  First, there is a large separation between the highest and next-highest occupied defect levels in the minority spin channel of $\mathrm{Al_{Mg}-Al_O}$, unlike the NV$^-$, where the $a_1$ and $b_1$ defect orbitals lie very close in energy to each other and to the valence band maximum. The increased energetic separation between the Kohn-Sham orbitals could lead to an increased separation between the low-lying excited states, thereby minimizing pJT effects and lowering the vibronic coupling, and leading to a lower value of $\Delta Q$. In fact, the energy separation between the vertical excitation energies of the first two $^3E$ states (calculated using TDDFT-DDH) is 1.64 eV for the $\mathrm{Al_{Mg}-Al_O}$ defect, compared to 0.67 eV for the NV$^-$ center (see  Fig. \ref{fig:es-prop}a). Thus, a strong vibronic coupling via the pJT effect is unlikely to occur in $\mathrm{Al_{Mg}-Al_O}$. Second, the occupied and unoccupied Kohn-Sham orbitals of the minority spin channel (which are involved in the excitation) are localized around the same Al atom in $\mathrm{Al_{Mg}-Al_O}$. This is unlike in NV$^-$, where the occupied defect orbitals localize near the Mg vacancy and the unoccupied defect orbitals localize on the N-O bond. Moreover, in the case of NV$^-$, the excitation is from the O 2$p$ orbitals of the Mg vacancy to the $\pi$-orbital on the N-O bond. Instead in $\mathrm{Al_{Mg}-Al_O}$, the excitation is between similar $\pi$-type orbitals on the same Al atom. Since a large change in the charge distribution of the defect electrons during an optical transition usually leads to a large HRF \cite{stoneham2001theory}, the comparatively low $\Delta Q$ of the $\mathrm{Al_{Mg}-Al_O}$ defect complex may originate, at least in part, from the localization of the orbitals involved in the transition around the same atom. 

Thus, spin defects in oxides comprising substitutional impurities  may lead to lower values of $\Delta Q$ and an increased DWF. These defects are expected to possess a large separation between Kohn-Sham defect levels (as compared to those with native defects such as cation vacancies) and result in the localization of the occupied and unoccupied defect states on the substitutional dopant itself. In this respect, $d$- or $f$-dopants might be promising, given that they likely substitute the cation site, and involve intra-shell $d$-$d$ or $f$-$f$ transitions, which might not experience a strong vibronic coupling due to similar charge density distributions of occupied and empty states. 

\section{Discussion}
In this work, we used the ADAQ high-throughput screening framework, along with hybrid DFT and TDDFT methods and QDET, to discover and characterize a promising NV$^-$ spin defect in MgO. The oxide host satisfies all the desirable criteria listed in the seminal paper by Weber et al. \cite{weber2010quantum}: it has a wide band gap, small spin-orbit coupling, it is available as high-quality thin films and bulk crystals, and both Mg and O have naturally occurring isotopes of zero nuclear spin and thus can be isotopically purified. Additionally, the recent study by Kanai et al. \cite{kanai2022generalized} predicted a single spin in MgO to have a coherence time ($T_2$) of 0.60 ms. This is slightly lower than the $T_2$ in prototypical hosts like diamond (0.89 ms) or SiC (1.1 ms) (primarily due to the larger concentration of nuclear spin isotopes of Mg). However, MgO could still be a promising host due to other advantages, particularly the ability to be grown as a high quality thin film via multiple routes \cite{plociennik2016optical, valanarasu2014microstructural}, and its use in a wide range of applications. These include MgO tunnel barriers in spintronics devices \cite{dieny2020opportunities}, MgO buffer layers in superconducting \cite{groves1999ion} and ferroelectric \cite{basit1998growth, murphy2004electronic} devices, and MgO heterogeneous catalyst support materials \cite{julkapli2016magnesium}. 

In addition to the favorable properties of the host, many properties of the NV$^-$ defect in MgO are promising as well. In particular, its spin triplet ground state provides a two-level system ($m_s$ = 0 and $m_s$ = $\pm$1), and the large ZFS can enable isolation of the two sublevels. The large ZFS and generally low hyperfine interaction parameters could be beneficial to increase the coherence time of the defect compared to the estimate obtained by considering a single electron spin in MgO. The singlet states between the $^3A_2$ ground and first $^3E$ excited state, as determined using TDDFT-DDH and QDET (see Fig. \ref{fig:es-prop}b), can act as shelving states, thereby enabling optical initialization and spin sublevel-dependent photo-luminescence. The efficiency of the initialization will depend on the radiative and non-radiative transition rates, which will be the subject of a future study. However, we note that the absence of crossing (or near-degeneracy) between the potential energy surfaces of the singlet and triplet states is an important indicator of the possibility of optical initialization and readout. The presence of singlet shelving states can also enable additional readout modes, for example spin-to-photocurrent (or spin-to-charge) readout. Another favorable feature of the NV$^-$ defect in MgO is its emission in the visible red-infrared regime, making the experimental detection of this defect relatively straightforward. It is unlikely that the emitted light will be reabsorbed, as one may expect  oxygen vacancies to be the additional most likely defect to be present in the oxide, and its photoexcitation and photoionization energies are around 5 eV \cite{vorwerk2023disentangling}. The broad emission line-width of the NV$^-$ center may allow filtering of any emitted light from underlying substrates, while still retaining a substantial portion of emitted light from the defect which can be detected. Moreover, recent work on Er$^{3+}$-doped MgO \cite{horvath2023strong} has demonstrated the integration of MgO with silicon nanophotonic cavities to achieve a cavity enhancement of over three orders of magnitude. This points to the promising possibility of using cavity enhancement of the optical transitions in the NV$^-$ center as well. Importantly, MgO can be doped with nitrogen, as demonstrated in previous experimental studies \cite{yang2010ferromagnetism,liu2011optical,li2013room,wang2021room,grob2012catalytic,li2013annealing}. The hyperfine interaction parameters of the NV$^-$ center are different from those of other nitrogen-related defects, e.g. the nitrogen interstitial and nitrogen substitutional on the oxygen site \cite{pesci2010nature}, indicating that the NV$^-$ center should be experimentally distinguishable.

The properties of the NV$^-$ spin defect computed here and the wide technological applicability of MgO thin films point at applications that could integrate `classical' spintronics and ferroelectrics with quantum information science. For example, classical information, e.g. magnetization direction in a magnetic tunnel junction (MTJ), with MgO as the tunnel barrier, can be mapped onto the spin defect, as it has been recently proposed by Jansen and Yuasa \cite{jansen2024high} in the case of a quantum dot. The NV$^-$ defect could also be used in conjunction with MgO MTJ-based nanomagnets to realize single qubit gates for quantum computing \cite{niknam2022quantum}. Similarly, it may be interesting to explore mapping of the polarization state of a ferroelectric thin film to control qubit interactions in MgO buffer layers \cite{levy2001quantum, liu2021coherent, george2013coherent}. Further, the properties of the NV$^-$ in MgO could  make it a viable quantum sensor. In fact, flexible and wearable spintronic devices based on MgO-barrier MTJs have been recently developed \cite{chen2017high}: combining the NV$^-$ spin defect with these devices could enable wearable quantum sensing. 

A key challenge with the promising defect identified in our study is the strong vibronic coupling, which leads to a large $\Delta Q$ and consequently large HRF and low DWF, leading to an undetectable ZPL. This likely makes the NV$^-$ center in MgO unsuitable for quantum communication; however, quantum sensing and computing could be viable applications. Our study revealed that strain could help increase the DWF, however, one would need to go to extremely large strains to achieve a detectable ZPL. Note that the DWF can be increased via the formation of point-planar defect complexes, like in SiC nanowires, where complexes formed between point defects and stacking faults were found to have high brightness and DWF $\approx$ 50\%. \cite{lee2021strong}. 
It is important to note that the large $\Delta Q$ and resulting low DWF found in our study at zero temperature represents the worst-case scenario, since all our analysis was done starting from a highly symmetric ground state structure. However,  the NV$^-$ defect in MgO has multiple nearly-degenerate local minima with lower symmetry. Given that zero-point vibrational energy in MgO is non-negligible (it leads to a lattice constant expansion by 0.5 \% and bandgap renormalization by 6 \% \cite{schimka2011improved, wing2021band}), it is likely that the zero-point vibrational energy will lead to symmetry breaking, which will reduce $\Delta Q$ and increase the DWF. Indeed, if we consider structures with broken symmetry in the ground state (as reported in our ADAQ search, see Supplementary Table 1), we find a decrease of the $\Delta Q$.

Importantly, our study was able to shed light onto the reasons behind the strong vibronic coupling found for the NV$^-$ in MgO. Our APES analysis showed that the strong vibronic coupling is due to the large pJT stabilization energy in combination with the low effective phonon frequency along the symmetry-breaking $e$-phonon modes. This finding gives insight into design principles that could be considered while exploring or screening new spin defects, especially for enhancing optical addressability and/or enabling quantum communication applications. In particular, the strength of the vibronic coupling due to the pJT effect increases when the energy separation between excited states decreases\cite{Bersuker_2006}. This behavior is often correlated with the energy separation between the single-particle Kohn-Sham defect levels. Thus, to lower the vibronic coupling (therefore lowering $\Delta Q$ and HRF), the localized mid-gap states (both occupied and unoccupied) must be well-separated from each other, which is what we observe in substitutional complexes such as the $\mathrm{Al_{Mg}-Al_O}$ defect in MgO (compare Fig. \ref{fig:engg-strat}b with Fig. \ref{fig:gs-prop}b for the NV$^-$). Moreover, the HRF increases when the charge distribution in the ground and excited state differ substantially in localization\cite{stoneham2001theory}, which is the case for the NV$^-$ defect in MgO (where the excitation is from the O $2p$ defect states around the magnesium vacancy to the $\pi$-orbital around the N-O bond). Instead, in VV$^0$ in SiC, NV$^-$ in diamond, and even in the $\mathrm{Al_{Mg}-Al_O}$ defect in MgO, the ground- and excited-state orbitals are localized in the same region of space. Thus, we expect spin defects in oxides comprising substitutional defects (including $d$- and $f$-dopants) to have lower vibronic coupling, leading to lower $\Delta Q$ and increased DWF. Moreover, our study shows that $\Delta Q$, the separation between Kohn-Sham levels, and the localization of the electronic states involved in the excitation are good descriptors for screening or designing spin defects. 

In sum, the computational protocol utilized in our study, combining high-throughput techniques and accurate electronic structure methods, helped identify and characterize a promising NV$^-$ defect in MgO. This new spin defect in a technologically mature material may be integrated into spintronic/ferroelectric applications to realize multifunctional electronics, including spintronic quantum logic gates and sensors. Overall, our results on the vibronic coupling of the NV$^-$ defect in MgO also contribute to the understanding of oxides as hosts for spin defects.

\section{Methods}
\subsection{ADAQ}

We used the ADAQ~\cite{ADAQ,adaq_info} software package and the high-throughput toolkit~\cite{armientoDatabaseDrivenHighThroughputCalculations2020} to create and screen defects in MgO, and the automatic screening workflow is presented in Ref.~\onlinecite{ADAQ}.

The MgO lattice parameter was optimized with the Perdew, Burke, and Erzenerhof (PBE) functional ~\cite{perdew1996generalized}, and we obtained 4.25 \AA. We used supercells with 512 atoms for all the defect calculations, and the Lany-Zunger charge correction~\cite{LanyZunger08} with a dielectric constant $\epsilon_0$ = 9.65.
In ADAQ, the calculations were performed with the Vienna Ab initio Simulation Package (VASP)~\cite{VASP,VASP2}, gamma compiled version 5.4.4, which uses the projector augmented wave (PAW)~\cite{PAW,Kresse99} pseudopotential (folder dated 2015-09-21).
We used a plane-wave energy cutoff of 600 eV and kinetic energy cutoff of 900 eV for the augmentation charges, a total energy criterion of $10^{-4}$ eV, and a structural minimization criterion of $5 \times 10^{-3}$ eV. Calculations were done at the $\Gamma$ point. All the 2917 defects screened can be found in the \href{https://defects.anyterial.se/search}{ADAQ database online}.

\subsection{Electronic structure calculations}
Once the NV$^-$- center was identified through the ADAQ procedure, we computed its ground state properties at the DFT level  using  Quantum ESPRESSO \cite{giannozzi2017advanced, giannozzi2009quantum, giannozzi2020quantum}. We used scalar-relativistic SG15 ONCV pseudopotentials (v. 1.2) \cite{hamann2013optimized, schlipf2015optimization} and the dielectric-dependent hybrid (DDH) \cite{skone2014self} exchange-correlation functional, with a plane-wave cutoff of 80 Ry. Convergence tests for the plane-wave cutoff energy are provided in Supplementary Note 10. The inverse of the experimental macroscopic dielectric constant $\epsilon_\infty$ was used to determine the fraction of exact exchange $\alpha$ ($\epsilon_\infty$ = 2.96 \cite{lide2004crc, skone2014self}, yielding $\alpha$ = $\epsilon_\infty^{-1}$ = 0.34) . Given the computational cost of hybrid functionals, we used supercells with 216 atoms and we fixed the lattice constant to the experimental value extrapolated to 0 K, including  the correction for zero-point anharmonic expansion effects \cite{skone2014self} ($a_0$ = 4.19 \AA). Note that this closely matches the DDH optimized lattice parameter (4.195 \AA) as obtained from an equation-of-state fit (see Supplementary Note 11). The supercell  Brillouin zone was sampled with the $\Gamma$ point. Forces were converged to 0.01 eV/\AA, and all calculations were spin-polarized. 

We calculated the formation energy of defect $D$ in charge state $q$ as \cite{freysoldt2014first}

\begin{equation}
\begin{split}
    E^f[D^q] = E_{tot}[D^q] - E_{tot}[bulk] - \sum_i n_i\mu_i \\
    + q(\varepsilon_f + E_{VBM}) + E_{corr}
\end{split}
\end{equation}

where $E^f[D^q]$ is the total energy of the supercell containing the defect $D$ in charge state $q$, $E_{tot}[bulk]$ is the total energy of the pristine supercell of the same size, $n_i$ is the number of atoms of species $i$ added to ($n_i>0$) or removed from ($n_i<0$) the supercell to create the defect, $\mu_i$ is the chemical potential of species $i$, $\varepsilon_f$ is the Fermi energy relative to the valence band maximum $E_{VBM}$, and $E_{corr}$ is a charge-dependent term that corrects for artificial electrostatic interactions arising due to the finite size of the supercell and was calculated using the Freysoldt-Neugebauer-Van de Walle correction scheme \cite{freysoldt2009fully} with the static dielectric constant for MgO $\epsilon_0=9.8$ taken from experiment \cite{subramanian1989dielectric}. Note that a comparison between the Lany-Zunger and Freysoldt-Neugebauer-Van de Walle correction schemes in the case of CaO \cite{davidsson2024discovery} yielded very similar and fully consistent results, and we expect the same for MgO. The chemical potentials of oxygen, nitrogen, and magnesium were calculated as \(\mu_O = \frac{1}{2}E_{DFT,O_2}\), \(\mu_N = \frac{1}{2}E_{DFT,N_2}\), \(\mu_{Mg} = E_{DFT, MgO} - \mu_O\). We used the HSE functional \cite{krukau2006influence} to obtain $E_{DFT,O_2}$ and $E_{DFT,N_2}$, as the binding energy and the bond length of the respective molecules calculated using the HSE functional matched more closely with experiments. Additional details can be found in Supplementary Note 12. A comparison of the defect formation energies computed at the PBE (the functional used in the ADAQ screening procedure) and DDH levels of theory is given in the Supplementary Information as well (Supplementary Note 13).

We computed the excited state properties, including the low-lying triplet and singlet states, and we optimized the geometry of the first excited triplet state using time-dependent DFT (TDDFT) under the Tamm-Dancoff approximation, the DDH functional and the WEST code \cite{jin2023excited, govoni2015large, yu2022gpu}. Due to the higher computational cost of TDDFT calculations, the plane-wave cutoff was reduced to 60 Ry. Forces were converged to 0.01 eV/\AA.

Vertical excitation energies and the resulting many-body electronic structure were also computed using the quantum defect embedding theory (QDET) with exact double counting corrections \cite{ma2021quantum, sheng2022green, vorwerk2022quantum}, using the WEST code \cite{govoni2015large, yu2022gpu}. The active space consisted of the localized defect orbitals ($b_1$, $a_1$, $e_x$, $e_y$) within the band gap as well as of several orbitals below the valence band maximum. We checked the convergence of the excitation energies by gradually increasing the number of orbitals below the valence band maximum (VBM) included in the active space, and found that convergence was achieved upon including orbitals up to 0.7 eV below the VBM. We note that in QDET with exact double counting correction, the environment is described using $G_0W_0$. In the WEST code, a separable form of $W_0$ is obtained using a technique called the projective eigendecomposition of the dielectric screening (PDEP), which avoids the inversion and storage of large dielectric matrices and enables the calculation of the self-energy matrix elements without explicitly evaluating empty states. The number of PDEP eigenpotentials used to obtain the separable form of $W_0$ was determined  after convergence tests of the excitation energies as a function of the number of eigenpotentials. The active space Hamiltonian was solved exactly using the full-configuration interaction method implemented in pySCF \cite{sun2018pyscf, sun2020recent}. Note that the electronic structure on which $G_0W_0$ and subsequent QDET calculations are based upon,  was obtained by non-polarized DFT calculations.

The zero field splitting tensor was calculated using the GPU-accelerated version of pyZFS \cite{ma2020pyzfs}, considering spin-spin interactions and using wavefunctions from a DFT calculation of a 4x4x4 (3x3x3) supercell with the PBE (DDH) exchange correlation functional. 

Hyperfine interaction parameters were calculated using the gauge-including projector-augmented wave (GIPAW) method with core polarization effects as implemented in the GIPAW module \cite{varini2013enhancement} of Quantum ESPRESSO. Wavefunctions from a DFT calculation of a 4x4x4 supercell with the PBE exchange correlation functional were used. 

\subsection{Configuration coordinate diagrams and emission spectra}
The one-dimensional configuration coordinate diagrams (CCD) were constructed by linearly interpolating the coordinates between the optimized $^3A_2$ ground state and first $^3E$ excited state structures and performing single point DFT and TDDFT calculations for each point along the ground and excited states respectively (using the DDH functional). The effective phonon frequencies in the ground and excited states ($\hbar\omega_g$ and $\hbar\omega_e$ respectively) were obtained by fitting a quadratic function around the points near the respective minima. The Huang-Rhys and Debye-Waller factors were computed using $\hbar\omega_g$, $\hbar\omega_e$, and $\Delta Q$, using Eqn.\ref{hrf}. 

Using the parameters obtained from the CCD, the normalized emission intensity $L(\hbar\omega)$ under the Franck-Condon approximation was calculated as

\begin{equation}\label{pl}
\begin{split}
    L(\hbar\omega) &= C\omega^3\sum_{i}\sum_{j}P_{ej}(T)|\langle\Theta_{ej}|\Theta_{gi}\rangle|^2 \\
    &\quad \times\delta(E_{ZPL}+E_{ej}-E_{gi}-\hbar\omega)
\end{split}
\end{equation}

Here $C$ is a normalization factor, which includes the electronic transition dipole moment and the refractive index of the material, $P_{ej}(T)$ is the thermal occupation factor of the excited state phonons with energy $E_{ej}$, $T$ is the temperature, $|\Theta_{gi}\rangle$ ($\langle\Theta_{ej}|$) is the $i$th ($j$th) harmonic nuclear wave function of the system in the ground (excited) state with vibrational energy $E_{gi}$ ($E_{ej}$), and $E_{ZPL}$ is the zero phonon line energy. The vibrational energies in the $n$th ground and excited states are given as $E_{gi}=n\hbar\omega_g$ and $E_{ej}=n\hbar\omega_e$ respectively. $E_{ZPL}$ is the energy difference between the optimized geometries of the excited and ground states. The Franck-Condon factors $|\langle\Theta_{ej}|\Theta_{gi}\rangle|^2$ were calculated using the recurrence method \cite{ruhoff1994recursion}. The $\delta$ function was replaced by a Gaussian with a broadening parameter of 25 meV. 

To quantify the coupling to the $e$-phonon modes, we built a configuration coordinate diagram between the averaged structure of the four equivalent excited state minima and the optimized geometry of the first excited state. Using the averaged structure as the reference one eliminates the contribution of the $a$-phonon modes, which are not Jahn-Teller (JT) or pseudo-Jahn Teller (pJT) active. The effective phonon frequencies and vibronic coupling constants were obtained by fitting an adiabatic potential energy surface (APES) to the CCD of the first excited state following the adiabatic-to-diabatic transformation method of Ref. \cite{viel2004effects, eisfeld2005higher} and including JT and pJT coupling terms up to the sixth order. 
\\

\section{Data availability}

For more information about the ADAQ database, see \url{https://httk.org/adaq/}.
Data that support the findings of this study will be made available at \url{https://qresp.org/} before publication.

\section{Competing interests}
The Authors declare no Competing Financial or Non-Financial Interests.

\section{Author contributions}
V.S.: calculations and analyses; J.D.: ADAQ methodology and high-throughput screening; Y.J.: TDDFT and PL methodology; G.G.: designed and supervised the research. V.S. and G.G. wrote the manuscript. All authors contributed to discussions and finalizing the manuscript. 

\begin{acknowledgments}
We thank Dr. F. Joseph Heremans, Prof. Shun Kanai, and Dr. Christian Vorwerk for many useful discussions. V.S acknowledges the support from the Maria Goeppert Mayer Named Fellowship, under the Laboratory Directed Research and Development (LDRD) funding from Argonne National Laboratory, provided by the Director, Office of Science, of the U.S. Department of Energy under Contract No. DE-AC02-06CH11357. J.D. acknowledges support from the Swedish e-science Research Centre (SeRC), the Knut and Alice Wallenberg Foundation through the WBSQD2 project (Grant No. 2018.0071), and the Swedish Research Council (VR) Grant No. 2022-00276. This work was supported by the Air Force Office of Scientific Research (AFOSR) through the CFIRE grant \# FA95502310667. This work used several codes, including the WEST code whose development is supported by MICCoM, which is part of the Computational Materials Sciences Program funded by the U.S. Department of Energy, Office of Science, Basic Energy Sciences, Materials Sciences, and Engineering Division through Argonne National Laboratory. This research used resources of the National Energy Research Scientific Computing Center (NERSC), a Department of Energy Office of Science User Facility using NERSC award DDR-ERCAP0029604, resources of the University of Chicago Research Computing Center, and resources of the Argonne Leadership Computing Facility, a U.S. Department of Energy (DOE) Office of Science user facility at Argonne National Laboratory, which is supported by the Office of Science of the U.S. DOE under Contract No. DE-AC02-06CH11357. The ADAQ computations were enabled by resources provided by the National Academic Infrastructure for Supercomputing in Sweden (NAISS) at the Swedish National Infrastructure for Computing (SNIC) at Tetralith, partially funded by the Swedish Research Council through grant agreement no. 2022-06725.
\end{acknowledgments}





\bibliography{main}

\end{document}


\preprint{APS/123-QED}

\title{Supplementary Information: An NV$^-$ center in magnesium oxide as a spin qubit for hybrid quantum technologies}%

\author{Vrindaa Somjit\footnote{\href{mailto:vsomjit@anl.gov}{vsomjit@anl.gov}}}
\affiliation{%
 Materials Science Division, Argonne National Laboratory, Lemont, IL 60439, USA
}%
\author{Joel Davidsson\footnote{\href{mailto:joel.davidsson@liu.se}{joel.davidsson@liu.se}}}
\affiliation{%
Department of Physics, Chemistry and Biology, Link\"oping University, SE-58183, Link\"oping, Sweden
}%
\author{Yu Jin\footnote{\href{mailto:jinyuchem@uchicago.edu}{jinyuchem@uchicago.edu}}}
\affiliation{
Pritzker School of Molecular Engineering and Department of Chemistry, University of Chicago, Chicago, IL 60637, USA
}%
\author{Giulia Galli\footnote{\href{mailto:gagalli@uchicago.edu}{gagalli@uchicago.edu}}}
\affiliation{Pritzker School of Molecular Engineering and Department of Chemistry, University of Chicago, Chicago, IL 60637, USA}
\affiliation{Materials Science Division, Argonne National Laboratory, Lemont, IL 60439, USA}

\date{November 19, 2024}

\maketitle



\section{List of all 40 defects from ADAQ}

Supplementary Tables \ref{tab:spin-1-NN} and \ref{tab:spin-1-NNN} list the 40 defects and defect complexes on the defect hull with spin triplet ground state and zero-phonon line as found by ADAQ. Note that the first 11 defects in Supplementary Table \ref{tab:spin-1-NN} are listed in the main text. The last five defects in Supplementary Table \ref{tab:spin-1-NN} consist of cations on the oxygen site or anions on the magnesium site, and therefore will likely be challenging to realize experimentally.

\begin{table}[H]
\caption{\justifying Isolated defects and defect complexes of first nearest-neighbors on the defect hull with spin triplet ground state and zero-phonon line}
\centering
\begin{tabular} {cc|r|rrr}
\hline
\hline
Defect & Defect & Charge & ZPL & TDM & $\Delta Q$ \\
Type & & & [eV] & [debye] & [amu$^{1/2}$\AA] \\
\hline
\multirow{4}{*}{$\mathrm{X_OVac_{Mg}}$}
& $\mathrm{B_OVac_{Mg}}$ & -1 &  0.5  &  1.93  &  1.45  \\
& $\mathrm{I_OVac_{Mg}}$  &  1   &  0.74  &  3.08  &  3.5  \\
& $\mathrm{Sb_OVac_{Mg}}$ &  -1 &  3.1  &  3.37  &  1.34   \\
& $\mathrm{Bi_OVac_{Mg}}$ &  -1 &  0.7  &  6.49  &  5.74 \\

\hline
\multirow{5}{*}{\shortstack{$\mathrm{X_{Mg}X_O}$}}
 & $\mathrm{B_{Mg}B_O}$ &  0  &  1.39  &  1.96  &  2.04   \\
 & $\mathrm{Al_{Mg}Al_O}$ &  0  &  0.68  &  1.35  &  1.0   \\
 & $\mathrm{Ga_{Mg}Ga_O}$ &  0    &  1.29  &  2.2  &  1.24   \\
 & $\mathrm{In_{Mg}In_O}$  &  0   &  0.98  &  2.7  &  1.2   \\
 & $\mathrm{Tl_{Mg}Tl_O}$   &  0   &  1.81  &  0.71  &  1.05   \\

\hline
\multirow{1}{*}{$\mathrm{X_O}$}
 & $\mathrm{Ga_O}$ &  -1   &  4.19  &  6.16  &  0.58  \\

\hline
\multirow{1}{*}{$\mathrm{Int_XVac_{Mg}}$}
& $\mathrm{Int_NVac_{Mg}}$  &  -1   &  2.19  &  0.66  &  1.95   \\

\hline
\multirow{2}{*}{$\mathrm{X_OInt_{O}}$}
& $\mathrm{Ge_OInt_{O}}$  &  0   &  2.5  &  4.77  &  3.45   \\
& $\mathrm{Sn_OInt_{O}}$  &  0   &  1.77  &  6.93  &  7.42   \\

\hline
\multirow{1}{*}{$\mathrm{I_XVac_{O}}$}
& $\mathrm{I_{Mg}Vac_{O}}$  &  1   &  0.74  &  3.08  &  3.5   \\

\hline
\multirow{2}{*}{$\mathrm{X_{Mg}Int_{X}}$}
& $\mathrm{Br_{Mg}Int_{Br}}$  &  0   &  0.91  &  2.49  &  1.99   \\
& $\mathrm{Cl_{Mg}Int_{Cl}}$  &  0   &  0.49  &  2.33  &  1.92   \\

 \hline
 \hline
\end{tabular}
\label{tab:spin-1-NN}
\end{table}
\begin{table}[H]
\caption{\justifying Isolated defects and defect complexes of second nearest-neighbors on the defect hull with spin triplet ground state and zero-phonon line}
\centering
\begin{tabular} {cc|r|rrr}
\hline
\hline
Defect & Defect & Charge & ZPL & TDM & $\Delta Q$ \\
Type & & & [eV] & [debye] & [amu$^{1/2}$\AA] \\

\hline
\multirow{5}{*}{\shortstack{$\mathrm{X_{Mg}Vac_{Mg}}$}}
 & $\mathrm{Rb_{Mg}Vac_{Mg}}$ &  -1  &  0.52  &  0.48  &  1.53   \\
 & $\mathrm{Ba_{Mg}Vac_{Mg}}$ &  0  &  0.47  &  2.14  &  1.18   \\
 & $\mathrm{H_{Mg}Vac_{Mg}}$ &  -1    &  2.43  &  1.85  &  1.44   \\
 & $\mathrm{O_{Mg}Vac_{Mg}}$  &  0   &  0.7  &  2.12  &  1.14   \\
 & $\mathrm{Cl_{Mg}Vac_{Mg}}$   &  -1   &  0.39  &  1.04  &  1.6   \\

\hline
\multirow{2}{*}{\shortstack{$\mathrm{X_{Mg}X_{Mg}}$}}
 & $\mathrm{H_{Mg}S_{Mg}}$ &  0  &  2.5  &  0.19  &  1.18   \\
 & $\mathrm{H_{Mg}S_{Mg}}$ &  0  &  0.87  &  5.26  &  1.53   \\

\hline
\multirow{2}{*}{\shortstack{$\mathrm{X_{Mg}X_{Mg}}$}}
 & $\mathrm{H_{Mg}S_{Mg}}$ &  0  &  2.5  &  0.19  &  1.18   \\
 & $\mathrm{H_{Mg}S_{Mg}}$ &  0  &  0.87  &  5.26  &  1.53   \\

\hline
\multirow{2}{*}{\shortstack{$\mathrm{X_{Mg}O_{Mg}}$}}
 & $\mathrm{I_{Mg}O_{Mg}}$ &  1  &  0.58  &  4.45  &  1.3   \\
 & $\mathrm{Bi_{Mg}O_{Mg}}$ &  -1  &  0.49  &  9.52  &  1.45   \\

\hline
\multirow{2}{*}{\shortstack{$\mathrm{X_{O}Mg_{O}}$}}
 & $\mathrm{Ga_{O}Mg_{O}}$ &  1  &  0.63  &  0  &  0.74   \\
 & $\mathrm{Li_{O}Mg_{O}}$ &  1  &  0.57  &  0.08  &  0.81   \\

\hline
\multirow{7}{*}{\shortstack{$\mathrm{X_{O}X_{O}}$}}
 & $\mathrm{N_{O}N_{O}}$ &  0  &  0.37  &  3.44  &  0.53   \\
 & $\mathrm{Ge_{O}Ge_{O}}$ &  0  &  0.63  &  0.23  &  1.09   \\
 & $\mathrm{Sn_{O}Sn_{O}}$ &  0  &  0.54  &  0.05  &  1.05   \\
 & $\mathrm{Si_{O}Si_{O}}$ &  0  &  0.68  &  0.17  &  0.93   \\
 & $\mathrm{Li_{O}Li_{O}}$ &  0  &  0.43  &  2.57  &  0.95   \\
 & $\mathrm{Mg_{O}Mg_{O}}$ &  0  &  0.2  &  7.38  &  0.82   \\
 & $\mathrm{Pb_{O}Pb_{O}}$ &  0  &  0.51  &  0.18  &  1.17   \\

\hline
\multirow{7}{*}{\shortstack{$\mathrm{X_{O}Vac_{O}}$}}
 & $\mathrm{Ca_{O}Vac_{O}}$ &  0  &  0.42  &  8.57  &  0.18   \\
 & $\mathrm{Sr_{O}Vac_{O}}$ &  0  &  0.39  &  7.86  &  1.06   \\
 & $\mathrm{Ga_{O}Vac_{O}}$ &  1  &  0.57  &  3.55  &  1.1   \\
 & $\mathrm{Al_{O}Vac_{O}}$ &  1  &  0.54  &  3.13  &  1.11   \\
 & $\mathrm{B_{O}Vac_{O}}$ &  1  &  0.99  &  2.39  &  0.96   \\
 & $\mathrm{Be_{O}Vac_{O}}$ &  0  &  0.38  &  6.22  &  0.98   \\
 & $\mathrm{Tl_{O}Vac_{O}}$ &  1  &  0.43  &  6.33  &  1.14   \\
 
 \hline
 \hline
\end{tabular}
\label{tab:spin-1-NNN}
\end{table}
\clearpage
\section{MgO band gap at different levels of theory and experiment}
Supplementary Table \ref{tab:bulk_props} tabulates the MgO band gap at different levels of theory and experiment.

\begin{table}[H]
    \caption{\justifying Computed band gaps of MgO at different levels of theory and experiment. Experimental band gap \cite{whited1973exciton} includes zero-point renormalization \cite{wing2021band}.}
    \centering
    \begin{tabular}{ccc}
    \hline
    \hline
         & Band gap (eV) \\
        \hline
        PBE & 4.47 \\
        DDH ($\alpha$ = 0.34) & 8.28 \\
        Experiment & 8.36 \cite{whited1973exciton, wing2021band} \\

        \hline
        \hline
    \end{tabular}
    \label{tab:bulk_props}
\end{table}
\clearpage
\section{Stability of the NV complex vs. $N_{Mg}$ substitutional defect}
To check if the nitrogen interstitial can recombine with the magnesium vacancy to form a nitrogen substitutional defect on the magnesium site ($N_{Mg}$), we carried out DFT calculations at the DDH level for the $N_{Mg}^q$ defect with different charge states $q$. We placed the nitrogen atom on the magnesium site and allowed all the ions to relax. We found that for $q$ = +1, 0, and -1, the system relaxes to form the NV center (i.e., the nitrogen atom prefers to bond with an oxygen atom, instead of staying at the substitutional site). Thus, $N_{Mg}^-$ is not stable and will always form the NV$^-$ defect. For $q$ = -2, $N_{Mg}^{2-}$ is higher in energy than the corresponding NV$^{2-}$ defect by 0.07 eV; whereas for $q$ = -3, $N_{Mg}^{3-}$ is higher in energy than the corresponding NV$^{3-}$ defect by 2.52 eV. The reason for this can be understood from a molecular orbital analysis.

\begin{figure}[H]
    \centering
    \includegraphics[scale=0.5]{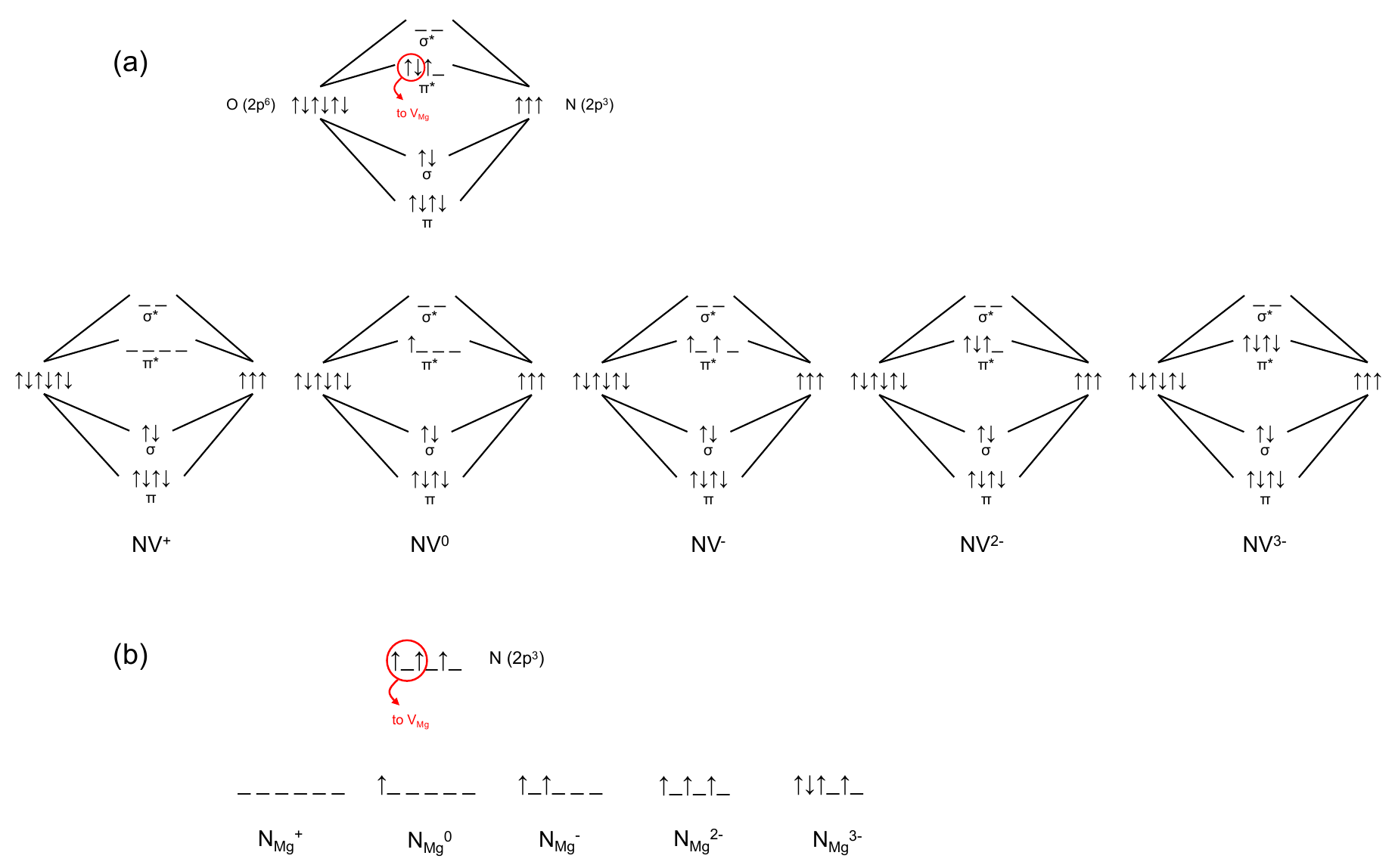}
    \caption{\justifying \textbf{a.} Molecular orbital diagram for the NV defect with different charge states. The bonding orbitals formed from the $p$-orbitals of O$^{2-}$ and N are shown. \textbf{b.} Orbital diagram from the $N_{Mg}$ defect with different charge states. The nitrogen $p$-orbitals are shown.}
    \label{fig:MO_diag_merged}
\end{figure}

Supplementary Figure \ref{fig:MO_diag_merged}a shows the molecular orbital diagrams for a nitrogen interstitial next to a magnesium vacancy (i.e., the NV defect), with different charge states $q$. When $q$ = 0, the nitrogen atom (2$s^2$2$p^3$) prefers to bond to the lattice O$^{2-}$ ion (2$s^2$2$p^6$) to form an N-O bond, resulting in the molecular orbital diagram shown in the top row of Supplementary Figure \ref{fig:MO_diag_merged}a. However, since the magnesium vacancy is an acceptor that can accept up to 2 electrons, 2 electrons from the $\pi^*$ orbital are donated to the nearby magnesium vacancy. This gives us the NV$^0$ defect, with one electron in the $\pi^*$ orbital. This also explains why the NV$^-$ defect has a triplet ground state- the additional electron will prefer to occupy the spin up state of the $\pi^*$ orbital following Hund’s rule. The different charge states of the NV$^q$ defect are shown at the bottom row of Supplementary Figure \ref{fig:MO_diag_merged}a.

Supplementary Figure \ref{fig:MO_diag_merged}b shows the orbital diagrams for a nitrogen substitutional on a magnesium site, with different charge states ($q$). This can be considered as a nitrogen atom on a magnesium vacancy. When $q$ = 0, the nitrogen atom has the electron configuration 2$s^2$2$p^3$. Once again, since the magnesium vacancy is an acceptor that can accept up to 2 electrons, 2 electrons from the $p$-orbital of N are donated to the magnesium vacancy. This gives us the $N_{Mg}^0$ defect, with one electron in the $p$-orbital. However, the NV$^0$ configuration is preferentially formed, due to energetic stabilization from the formation of the N-O bond. This N-O bond formation is what promotes the formation of NV$^+$, NV$^0$, and NV$^-$, instead of the corresponding $N_{Mg}^+$, $N_{Mg}^0$, and $N_{Mg}^-$. However, $N_{Mg}^{2-}$ is nearly as stable as NV$^{2-}$, likely due to the stability of the half-filled $p$-orbital on N in $N_{Mg}^{2-}$. For the same reason, NV$^{3-}$ is much more stable than $N_{Mg}^{3-}$, due to the fully-filled $\pi^*$ molecular orbital in NV$^{3-}$.
\clearpage
\section{Migration barrier calculations}
\begin{figure}[H]
    \centering
    \includegraphics[scale=0.5]{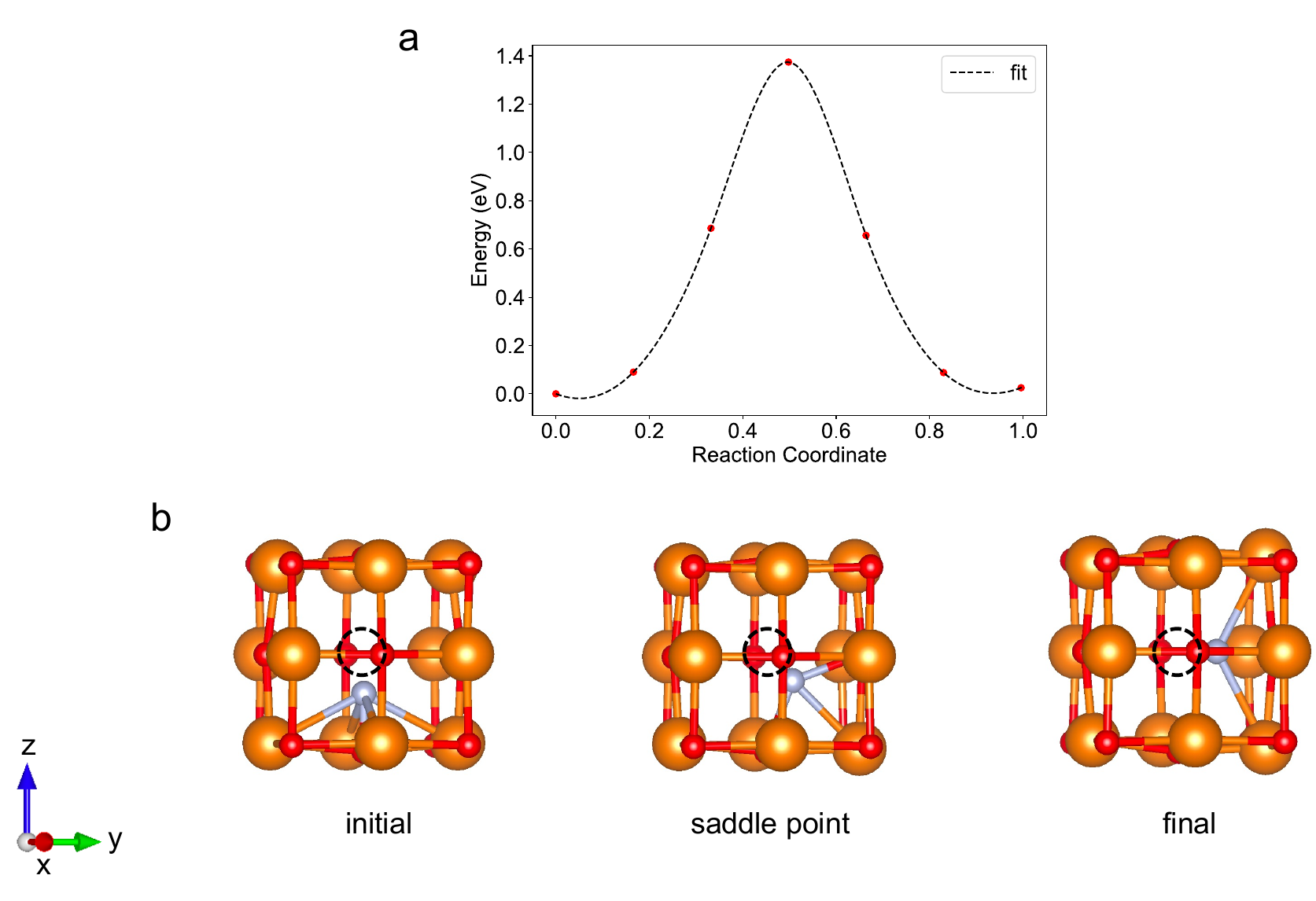}
    \caption{\justifying \textbf{a.} Minimum energy pathway for the migration of the nitrogen interstitial from one side of the magnesium vacancy to another. \textbf{b.} Initial, saddle point, and final configurations.}
    \label{fig:neb_ddh}
\end{figure}

We computed the energy barrier for the nitrogen interstitial to migrate from one interstitial site to another interstitial site in the vicinity of the magnesium vacancy using the climbing image nudged elastic band method \cite{henkelman2000climbing} (216 atom supercell, plane-wave cutoff of 80 Ry, $\Gamma$-point sampling, DDH functional). Supplementary Figure \ref{fig:neb_ddh}a plots the minimum energy pathway for nitrogen migration, and Supplementary Figure \ref{fig:neb_ddh}b shows the initial, saddle point, and final configurations. The computed migration barrier is 1.37 eV. Thus, once formed, this defect is expected to be stable against annealing.

\clearpage
\section{Hyperfine interaction parameters of nitrogen defects in MgO}
Supplementary Table \ref{tab:hfi-comp} compares the hyperfine interaction parameters of various nitrogen defects in MgO. Note how the reported hyperfine parameters for substitutional nitrogen on the oxygen site and interstitial nitrogen are different from our calculated values for the NV$^-$ defect. 
\begin{table}[h]
    \caption{\justifying Computed principal  values of the hyperfine tensors for different nitrogen defects in MgO at the PBE level of theory (in G).}
    \centering
    \begin{tabular}{cccc}
    \hline
    \hline
        Atom & A$_{xx}$ & A$_{yy}$ & A$_{zz}$ \\
        \hline
        N atom in NV$^-$ defect\footnote{This work} & 15.55 & 15.55 & -31.10 \\
        O atom bonded to N in NV$^-$ defect$\mathrm{^a}$ & -9.94 & -9.95 & 19.89 \\
        N substitution on O site\footnote{Ref. \cite{pesci2010nature}} & -11.9 & -11.9 & 23.8 \\
        N interstitial$\mathrm{^b}$ & -11.8 & -11.3 & 23.2 \\
        O atom bonded to N interstitial$\mathrm{^b}$ & 15.2 & 15.3 & -30.5 \\
        \hline
        \hline
    \end{tabular}
    \label{tab:hfi-comp}
\end{table}

\clearpage
\section{Construction of the configuration coordinate diagram}

The character table for the $C_{4v}$ point group is given in Supplementary Table \ref{tab:c4v}. The $E$ electronic states can couple linearly to $a_1$, $a_2$, $b_1$, $b_2$ phonon modes, and to $e$ phonon modes via higher (even) order coupling. However, symmetry can be broken only by the $b_1$, $b_2$, and $e$ phonon modes. The optimized geometry of the $^3E$ excited state belongs to the $C_1$ point group, thus, $e$ phonon modes must be predominant in reducing the symmetry from the $C_{4v}$ point group to $C_1$. 

\begin{table*}[ht]
\caption{Character table for $C_{4v}$ point group}
\centering
\begin{tabular}{c|c|c|c|c|c|c|c|c}
\hline
\hline
$\mathrm{C_{4v}}$ & E & 2$\mathrm{C_4(z)}$ & $\mathrm{C_2}$ & 2$\mathrm{\sigma_v}$ & 2$\mathrm{\sigma_d}$ & linear functions, rotations & quadratic functions & cubic functions \\ \hline
$\mathrm{A_1}$ & +1 & +1 & +1 & +1 & +1 & z & $x^2 + y^2, z^2$ & $z^3, z(x^2 + y^2)$ \\ \hline
$\mathrm{A_2}$ & +1 & +1 & +1 & -1 & -1 & $R_z$ & - & - \\ \hline
$\mathrm{B_1}$ & +1 & -1 & +1 & +1 & -1 & - & $x^2 - y^2$ & $z(x^2 - y^2)$ \\ \hline
$\mathrm{B_2}$ & +1 & -1 & +1 & -1 & +1 & - & $xy$ & $xyz$ \\ \hline
$\mathrm{E}$ & +2 & 0 & -2 & 0 & 0 & $(x, y) (R_x, R_y)$ & $(xz, yz)$ & $(xz^2, yz^2) (xy^2, x^2y) (x^3, y^3)$ \\ \hline
\hline
\end{tabular}
\label{tab:c4v}
\end{table*}

There are four equivalent minima with $C_1$ point symmetry. To verify that the $e$ phonon modes are indeed predominantly involved in symmetry-breaking in the first excited $^3E$ state, we calculated the Mg, O, and N displacements between the average structure of the four equivalent minima and the optimized excited state structure. The average structure has contributions from only the $a_1$-phonon modes (as contributions from the symmetry-breaking modes cancel out, since they are equal in magnitude but opposite in direction for the four equivalent minima); therefore, the displacement between the average structure and the optimized excited state structure reveals the predominant symmetry-breaking mode.

\begin{table}[H]
\caption{Displacement components of the N atom, and O and Mg atoms nearest neighbor to the N atoms along the x-, y-, and z- directions. Unit is \AA.}
\centering
\begin{tabularx}{\textwidth}{c|>{\centering\arraybackslash}X|>{\centering\arraybackslash}X|>{\centering\arraybackslash}X|>{\centering\arraybackslash}X|>{\centering\arraybackslash}X|>{\centering\arraybackslash}X}
\hline
\hline
   & \multicolumn{3}{|c|}{Between average and ground state structures} & \multicolumn{3}{|c}{Between excited state and average structures} \\ \hline
 & $\mathrm{x_{es}-x_{gs}}$ & $\mathrm{y_{es}-y_{gs}}$ & $\mathrm{z_{es}-z_{gs}}$ & $\mathrm{x_{es}-x_{gs}}$ & $\mathrm{y_{es}-y_{gs}}$ & $\mathrm{z_{es}-z_{gs}}$ \\ \hline
$\mathrm{Mg_1}$ & 0.04 & 3.35e-05 & 7.98e-03 & -0.08 & 5.82e-06 & 1.11e-03 \\ \hline
$\mathrm{Mg_2}$ & -3.34e-05 & 0.04 & 7.98e-03 & 9.54e-03 & 0.10 & 0.05 \\ \hline
$\mathrm{Mg_3}$ & -0.04 & -3.34e-05 & 7.98e-03 & 0.13 & -2.67e-04 & -0.10 \\ \hline
$\mathrm{Mg_4}$ & 3.35e-05 & -0.04 & 7.98e-03 & 9.27e-03 & -0.10 & 0.05 \\ \hline
N & 0 & 0 & -0.39 & 0.51 & 1.70e-04 & -2.92e-07 \\ \hline
O & 0 & 0 & -0.13 & -0.34 & -4.15e-05 & -1.86e-07 \\ \hline
\hline
\end{tabularx}
\label{tab:disp}
\end{table}

\begin{figure}[H]
    \centering
    \includegraphics[width=1\linewidth]{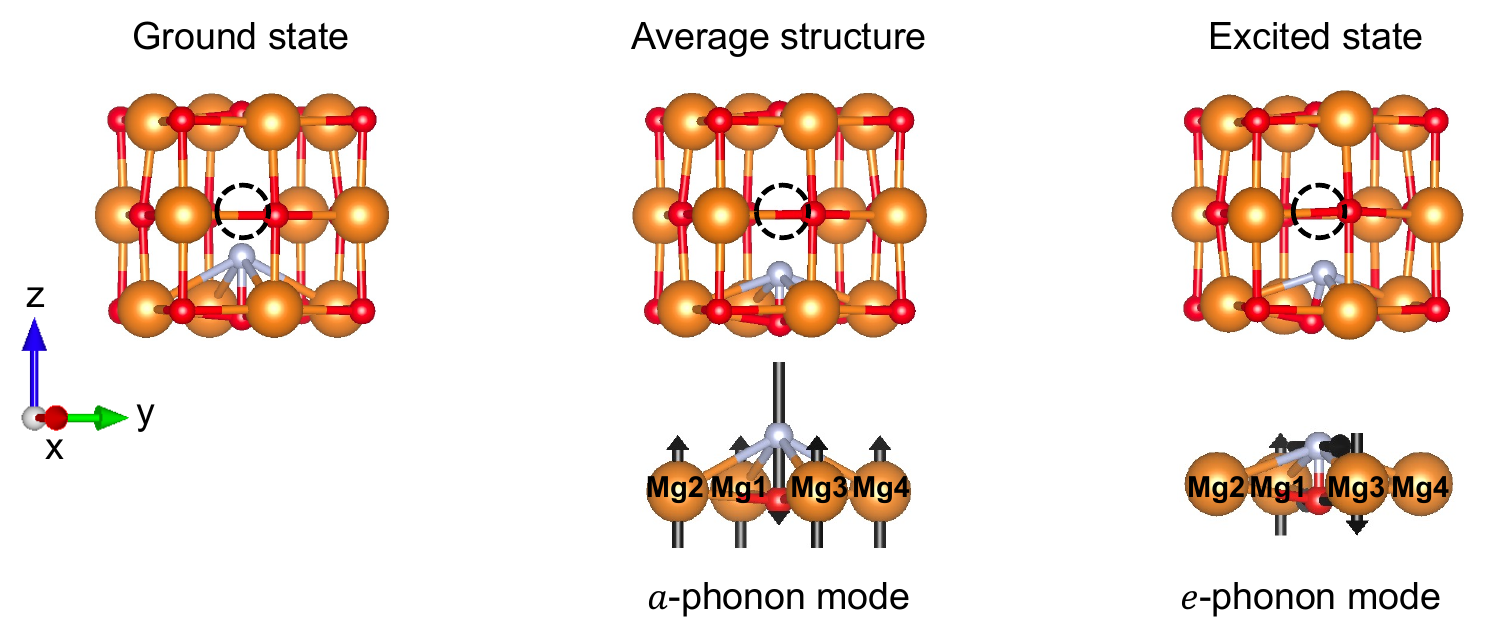}
    \caption{\justifying \textbf{Top row}: Optimized geometry of the ground state structure, average structure of the four equivalent excited state minima, optimized geometry of the $^3E$ excited state (i.e. one of the four equivalent minima). \textbf{Bottom row}: predominant phonon modes involved in generating the average structure (from the ground state geometry) and the excited state structure (from the average structure). The displacement vectors for $\mathrm{Mg_1}$, $\mathrm{Mg_2}$, $\mathrm{Mg_3}$, $\mathrm{Mg_4}$ of the $a$-phonon modes are shown only along the z-direction and are scaled up by 38 times for clarity. Similarly, the displacement vectors for $\mathrm{Mg_1}$ and $\mathrm{Mg_3}$ of the e-phonon modes are scaled by 300 and 3 times respectively.}
    \label{fig:phonons}
\end{figure}

Supplementary Table \ref{tab:disp} lists the displacement values between the average structure and the ground state structure, and the excited state structure and the average structure, for the N atom, and the O and Mg atoms nearest neighbor to the N atom. We find that $a_1$-phonon modes are predominant in generating the average structure (from the ground state structure), and $e$-modes are predominant in generating the optimized geometry in the excited state (from the average structure). We compared the observed phonon modes to the vibrational modes reported for the $\mathrm{XeOF_4}$ molecule \cite{claassen1963vibrational}, which has $C_{4v}$ symmetry and a geometry similar to that of the NV$^-$ defect center. The $a_1$-phonon mode is akin to the `out of plane bend' mode, and the $e$-phonon mode is the `bend' mode. Supplementary Figure \ref{fig:phonons} shows the ground state, average, and excited state structures, and the predominant phonon modes involved in each structure. 

Thus, to calculate the vibronic coupling of the $E$ electronic states with the $e$ phonon modes, we take the average structure as the reference configuration for our configuration coordinate diagram (CCD), and interpolate points between this new reference configuration and one of the minima. This eliminates the influence of $a_1$-phonon modes (which are not Jahn-Teller or pseudo Jahn-Teller active) and gives the coupling of the $E$-states exclusively to the $e$-modes. 

\clearpage
\section{Construction of the adiabatic potential energy surface}
To construct the adiabatic potential energy surface (APES) of the $^3E$ electronic state along the $e$-phonon mode and obtain the effective phonon frequencies and vibronic coupling constants, we followed the method proposed in \cite{viel2004effects} and \cite{eisfeld2005higher}. We extended their method (which was developed for a $C_{3v}$ $E \otimes e$ Jahn-Teller system \cite{viel2004effects} and a $C_{3v}$ $(A+E) \otimes e$ pseudo Jahn-Teller system \cite{eisfeld2005higher}) to our $C_{4v}$ $(E+E) \otimes e$ system. The method has been extensively described in previous works, and we only provide a short summary  here. 

Each component of the doubly-degenerate $^3E$ state can interact via the Jahn-Teller effect, and the closely-spaced $^3E$-states (shown in Fig. 3b and Fig. 4b in the main text) can interact via the pseudo Jahn-Teller effect. Thus, we build an effective Hamiltonian $\hat{H}$ that accounts for both of these effects up to the sixth order. To do so, we expand the elements of $\hat{H}$ in a Taylor series up to the sixth order along the $e_x$ and $e_y$ phonon modes (i.e., along $x$ and $y$ directions in real space). To identify the non-vanishing terms of the expansion coefficients, we use the fact that $\hat{H}$ must be invariant under the symmetry operations $\hat{S}$ of the $C_{4v}$ point group, since \([\hat{H},\hat{S}] = 0\). The symmetry operations $\hat{S}$ are given in Supplementary Table \ref{tab:c4v}, namely, $\hat{C_4}$ (rotation by $\pi/2$), $\hat{C_2}$ (rotation by $\pi$), $\hat{\sigma_v}$ (reflection on mirror plane along $C_4$ axis), and $\hat{\sigma_d}$ (reflection along diagonal plane). 

By doing this, we get the matrix 

\begin{widetext}
\renewcommand{\arraystretch}{2} 
\[
\hat{H} = \frac{1}{2}\sum_{n=0}^{6}\frac{1}{n!}\begin{Bmatrix}
\begin{pmatrix}
V_1^{(n)}+W_{JT,1}^{(n)} & M_{JT,1}^{(n)} & Z_{PJT,1}^{(n)}+P_{PJT,1}^{(n)} & Z_{PJT,2}^{(n)}+P_{PJT,2}^{(n)} \\
M_{JT,1}^{(n)}   & V_1^{(n)}-W_{JT,1}^{(n)} & -Z_{PJT,2}^{(n)}+P_{PJT,2}^{(n)} & Z_{PJT,1}^{(n)}-P_{PJT,1}^{(n)} \\
 Z_{PJT,1}^{(n)}+P_{PJT,1}^{(n)}   & -Z_{PJT,2}^{(n)}+P_{PJT,2}^{(n)} & V_2^{(n)}+W_{JT,2}^{(n)} & M_{JT,2}^{(n)} \\
Z_{PJT,2}^{(n)}+P_{PJT,2}^{(n)}   & Z_{PJT,1}^{(n)}-P_{PJT,1}^{(n)} & M_{JT,2}^{(n)} & V_2^{(n)}-W_{JT,2}^{(n)} \\
\end{pmatrix}
\end{Bmatrix}
\]
\renewcommand{\arraystretch}{1} 
\end{widetext}

The explicit equations are as follows

\begin{equation}
    V_1^{(0)} = a_1^{(0)}
\end{equation}
\begin{equation}
    V_1^{(2)} = a_1^{(2)}[x^2+y^2]
\end{equation}	
\begin{equation}
    \begin{split}
    V_1^{(4)} &= a_1^{(4)}[2x^4 - 12x^2y^2 + 2y^4] \\
    &\quad + a_2^{(4)}[x^4 + 2x^2y^2 + y^4]
    \end{split}
\end{equation}	
\begin{equation}
    \begin{split}
    V_1^{(6)} &= a_1^{(6)}[2x^6 - 10x^4y^2 - 10x^2y^4 + 2y^6] \\
    &\quad + a_2^{(6)}[x^6 + 3x^4y^2 + 3x^2y^4 + y^6]
    \end{split}
\end{equation}

\begin{equation}
    V_2^{(0)} = b_1^{(0)}
\end{equation}
\begin{equation}
    V_2^{(2)} = b_1^{(2)}[x^2+y^2]
\end{equation}	
\begin{equation}
    \begin{split}
    V_2^{(4)} &= b_1^{(4)}[2x^4 - 12x^2y^2 + 2y^4] \\
    &\quad + b_2^{(4)}[x^4 + 2x^2y^2 + y^4]
    \end{split}
\end{equation}	
\begin{equation}
    \begin{split}
    V_2^{(6)} &= b_1^{(6)}[2x^6 - 10x^4y^2 - 10x^2y^4 + 2y^6] \\
    &\quad + b_2^{(6)}[x^6 + 3x^4y^2 + 3x^2y^4 + y^6]
    \end{split}
\end{equation}

\begin{equation}
    W_{JT,1}^{(2)} = \lambda_1^{(2}[2x^2 - 2y^2]
\end{equation}
\begin{equation}
    W_{JT,1}^{(4)} = \lambda_1^{(4)}[2x^4 - 2y^4]
\end{equation}	
\begin{equation}
    \begin{split}
    W_{JT,1}^{(6)} &= \lambda_1^{(6)}[2x^6 - 30x^4y^2 + 30y^4x^2 - 2y^6] \\
    &\quad + \lambda_2^{(6)}[2x^6 + 2x^4y^2 - 2y^4x^2 - 2y^6] 
    \end{split}
\end{equation}	

\begin{equation}
    W_{JT,2}^{(2)} = \tau_1^{(2)}[2x^2 - 2y^2]
\end{equation}
\begin{equation}
    W_{JT,2}^{(4)} = \tau_1^{(4)}[2x^4 - 2y^4]
\end{equation}	
\begin{equation}
    \begin{split}
    W_{JT,2}^{(6)} &= \tau_1^{(6)}[2x^6 - 30x^4y^2 + 30y^4x^2 - 2y^6] \\
    &\quad + \tau_2^{(6)}[2x^6 + 2x^4y^2 - 2y^4x^2 - 2y^6] 
    \end{split}
\end{equation}	

\begin{equation}
    M_{JT,1}^{(2)} = \sigma_1^{(2)}[4xy]
\end{equation}
\begin{equation}
    M_{JT,1}^{(4)} = \sigma_1^{(4)}[4xy^3 + 4x^3y]
\end{equation}	
\begin{equation}
    \begin{split}
    M_{JT,1}^{(6)} &= \sigma_1^{(6)}[12x^5y - 40x^3y^3 + 12xy^5] \\
    &\quad + \sigma_2^{(6)}[4x^5y + 8x^3y^3 + 4xy^5] 
    \end{split}
\end{equation}	

\begin{equation}
    M_{JT,2}^{(2)} = \theta_1^{(2)}[4xy]
\end{equation}
\begin{equation}
    M_{JT,2}^{(4)} = \theta_1^{(4)}[4xy^3 + 4x^3y]
\end{equation}	
\begin{equation}
    \begin{split}
    M_{JT,2}^{(6)} &= \theta_1^{(6)}[12x^5y - 40x^3y^3 + 12xy^5] \\
    &\quad + \theta_2^{(6)}[4x^5y + 8x^3y^3 + 4xy^5] 
    \end{split}
\end{equation}	

\begin{equation}
    Z_{PJT,1}^{(0)} = c_1^{(0)}
\end{equation}
\begin{equation}
    Z_{PJT,1}^{(2)} = c_1^{(2)}[x^2 + y^2]
\end{equation}	
\begin{equation}
    \begin{split}
    Z_{PJT,1}^{(4)} &= c_1^{(4)}[2x^4 - 12x^2y^2 + 2y^4] \\
    &\quad + c_2^{(4)}[x^4 + 2x^2y^2 + y^4] 
    \end{split}
\end{equation}	
\begin{equation}
    \begin{split}
    Z_{PJT,1}^{(6)} &= c_1^{(6)}[2x^6 - 10x^4y^2 - 10x^2y^4 + 2y^6] \\
    &\quad + c_2^{(6)}[x^6 + 3x^4y^2 + 3x^2y^4 + y^6] 
    \end{split}
\end{equation}	

\begin{equation}
    P_{PJT,1}^{(2)} = \eta_1^{(2)}[2x^2 - 2y^2]
\end{equation}	
\begin{equation}
    P_{PJT,1}^{(4)} = \eta_1^{(4)}[2x^4 - 2y^4] 
\end{equation}	
\begin{equation}
    \begin{split}
    P_{PJT,1}^{(6)} &= \eta_1^{(6)}[2x^6 - 30x^4y^2 + 30x^2y^4 - 2y^6] \\
    &\quad + \eta_2^{(6)}[2x^6 + 2x^4y^2 - 2x^2y^4 - 2y^6] 
    \end{split}
\end{equation}	

\begin{equation}
    Z_{PJT,2}^{(2)} = \mu_1^{(2)}[4xy]
\end{equation}	
\begin{equation}
    \begin{split}
    Z_{PJT,2}^{(4)} &= \mu_1^{(4)}[4xy^3 + 4x^3y] 
    \end{split}
\end{equation}	
\begin{equation}
    \begin{split}
    Z_{PJT,2}^{(6)} &= \mu_1^{(6)}[12x^5y - 40x^3y^3 + 12xy^5] \\
    &\quad + \mu_2^{(6)}[4x^5y + 8x^3y^3 + 4xy^5] 
    \end{split}
\end{equation}	

\begin{equation}
    \begin{split}
    P_{PJT,2}^{(4)} &= d_1^{(4)}[8x^3y - 8xy^3] 
    \end{split}
\end{equation}	
\begin{equation}
    \begin{split}
    P_{PJT,2}^{(6)} &= d_1^{(6)}[8x^5y - 8xy^5] 
    \end{split}
\end{equation}	

The vibronic coupling parameters were determined by fitting the eigenvalues of $\hat{H}$ to the TDDFT-DDH single-point energies of the first $^3E$ excited state along the $e_x$ and $e_y$ phonon modes (i.e., to the CCD as obtained in the previous section). Note that we tried fitting the APES with expansions of only up to two- and four-orders, but the fit was poor. The fits with two-, four-, and six- orders are given in Supplementary Figure \ref{fig:fits}. Moreover, it is indeed expected that the strong vibronic coupling observed in our system requires higher order coupling parameters to fit the APES. 

\begin{figure}[H]
    \centering
    \includegraphics[width=0.75\linewidth]{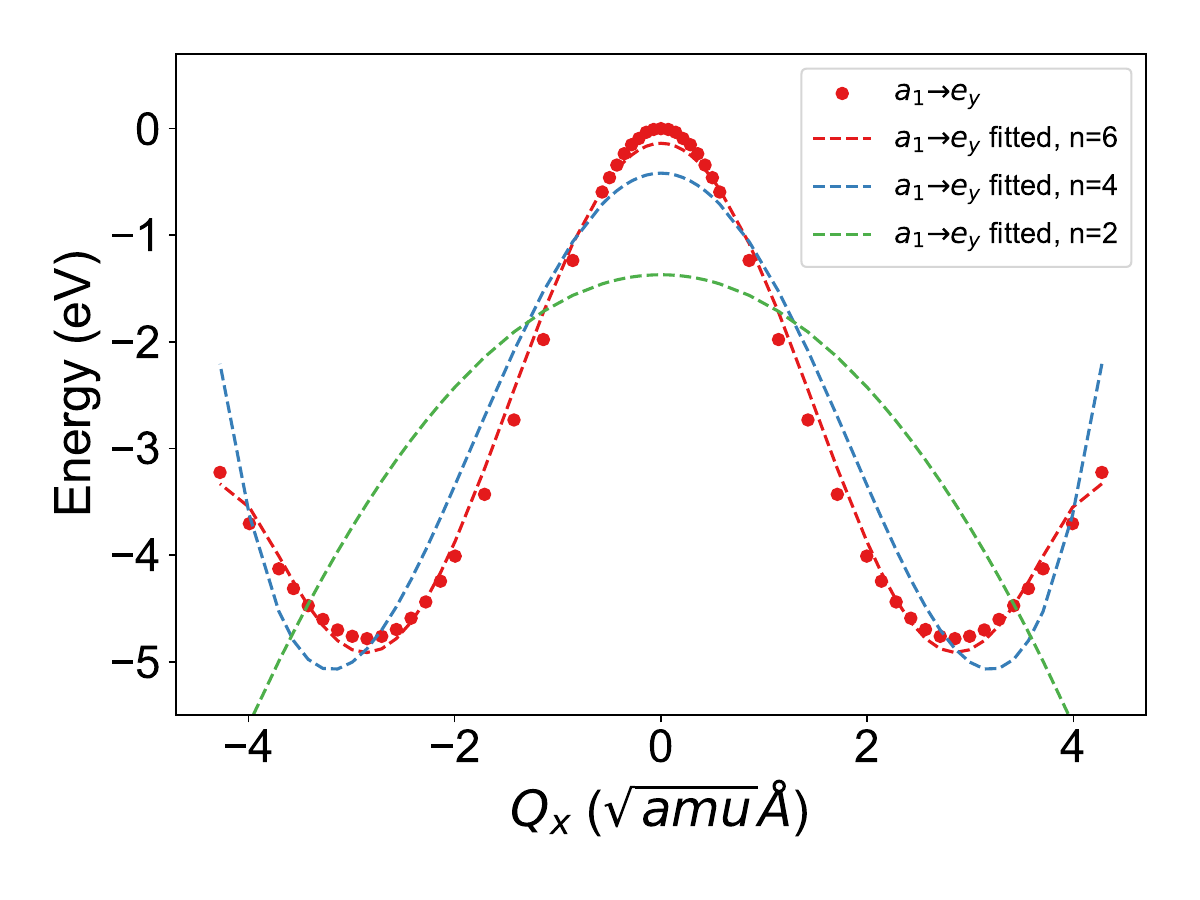}
    \caption{\justifying Adiabatic potential energy surface of first $^3E$-excited state along the $e_x$-phonon mode obtained with expansions of two-, four- and six-orders. The markers denote energies obtained at each configuration using single-point DDH (TDDFT-DDH) calculations.}
    \label{fig:fits}
\end{figure}

The fitting procedure up to the sixth order is challenging, due to the large number of unknown parameters. Thus, while solving for the coupling parameters, we made several approximations:
\begin{enumerate}
    \item We assume $a_{1,2}^{(n)}$ = $b_{1,2}^{(n)}$, for $n>0$ (i.e. non-coupling coefficients for different excited $^3E$ states are the same).
    \item We found that including Jahn-Teller terms did not reduce the fitting error; i.e. in this system, the pseudo Jahn-Teller effect is dominant. Thus, we set $\lambda_{1,2}^{(n)}$ = $\tau_{1,2}^{(n)}$ = $\sigma_{1,2}^{(n)}$ = $\theta_{1,2}^{(n)}$ = 0. 
    \item We set $\mu_{1,2}^{(n)}$, $d_{1,2}^{(n)}$ = 0, since these are coefficients of terms involving products of $x$ and $y$, which will be 0 since we project the APES along either the $x$-axis ($y$ = 0) or the $y$-axis ($x$ = 0). 
    \item We assume only 2 excited states interact (i.e.we consider only a two-level pseudo Jahn-Teller effect instead of a multi-level pJT effect), and we fit only to the energies of the first excited state. Accounting for multi-level pJT would further increase the matrix size, leading to a larger number of unknown parameters.
\end{enumerate}
These approximations helped reduce the number of unknown parameters while still retaining a realistic physical representation of the problem. Moreover, given the complicated loss function that needed to be minimized, we tested the influence of different initial parameters. We also tested for overfitting by fitting the curves using 80\% of the data points, and calculating the error on the remaining 20\% data points. We consistently found the effective phonon frequency to be around 20 meV and vibronic coupling parameters to be 20-500 meV. The vibronic coupling coefficients are given in Supplementary Table \ref{tab:coeff}.

\begin{table}[H]
\centering
\caption{Non-coupling and pseudo Jahn Teller (pJT) coupling coefficients (meV)}
\begin{tabular}{c|c|c|c}
\hline
\hline
\multirow{2}{*}[4pt]{\textbf{Non-coupling}} & \multirow{2}{*}[4pt]{\textbf{Value}} & \multirow{2}{*}[4pt]{\textbf{pJT coupling}} & \multirow{2}{*}[4pt]{\textbf{Value}} \\ 
 \textbf{coefficient} & \textbf{(meV)} & \textbf{coefficient} & \textbf{(meV)} \\ \hline
$a_1^{(0)}$ & -274.62 & $c_1^{(2)}$ & 17.71 \\ \hline
$a_1^{(2)}$ & 25.81 & $c_1^{(4)}$ & 6.76 \\ \hline
$a_1^{(4)}$ & 9.50 & $c_2^{(4)}$ & -16.47 \\ \hline
$a_2^{(4)}$ & -10.99 & $c_1^{(6)}$ & 7.61 \\ \hline
$a_1^{(6)}$ & 29.62 & $c_2^{(6)}$ & -14.77 \\ \hline
$a_2^{(6)}$ & 29.24 & $\eta_1^{(2)}$ & 448.57\\ \hline
$b_1^{(0)}$ & -274.61 & $\eta_1^{(4)}$ & -70.93 \\ \hline
 &  & $\eta_1^{(6)}$ & 26.68 \\ \hline
 &  & $\eta_2^{(6)}$ & 26.68 \\ \hline
\hline
\end{tabular}
\label{tab:coeff}
\end{table}

\clearpage
\section{Ground state phonon modes}
We calculated the phonon modes for the NV$^-$ defect at the ground state using the frozen phonon approach as implemented in the Phonopy package \cite{togo2023first, togo2023implementation}. The displaced configurations of the 216-atom supercell were generated using a displacement of 0.01 \AA from the ground state, and the self-consistent calculations were carried out using the DDH functional (with a plane-wave cutoff of 80 Ry and $\Gamma$-point sampling).

\begin{figure}[H]
    \centering
    \includegraphics[scale=0.5]{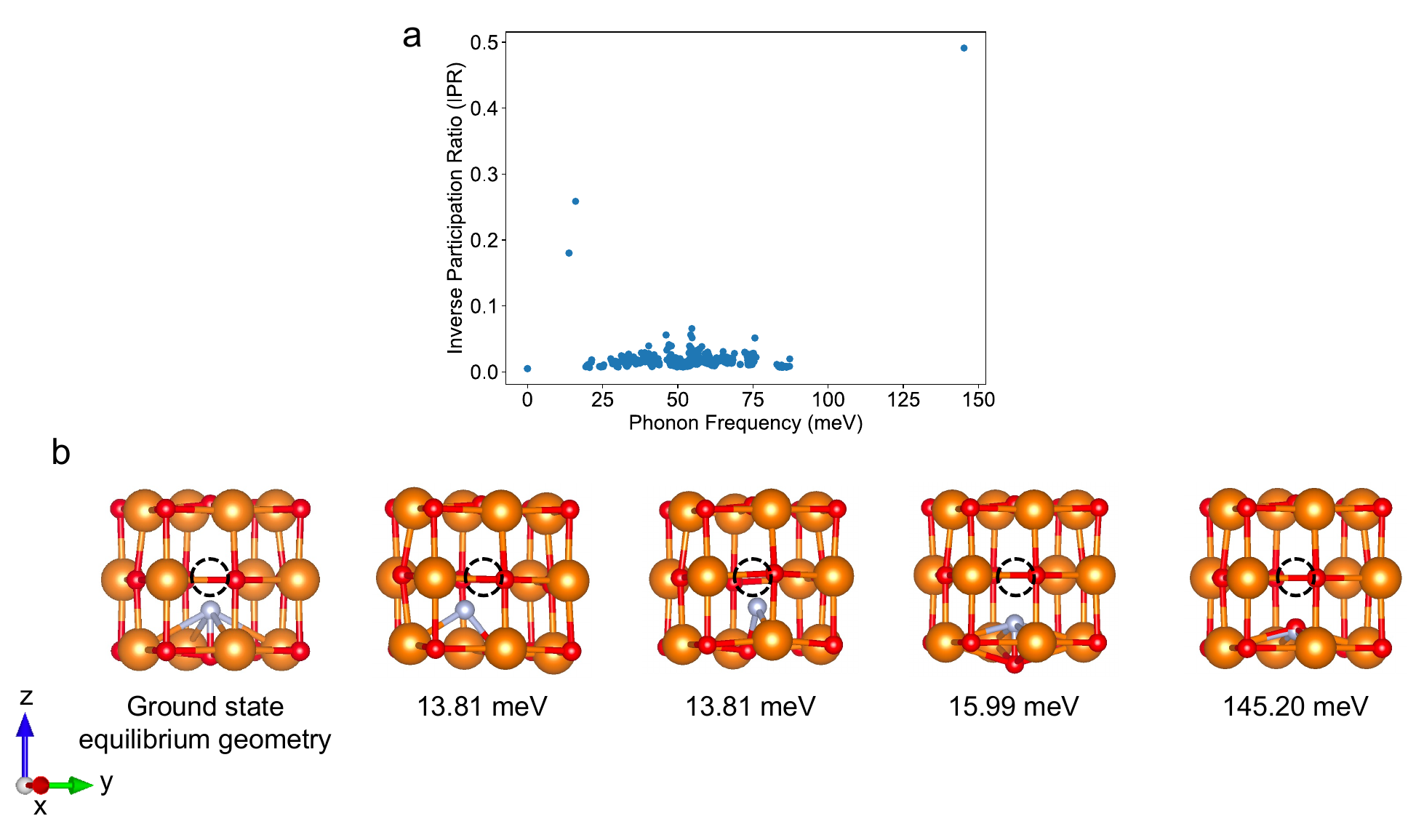}
    \caption{\justifying \textbf{a.} Inverse participation ratio of phonon eigenvectors calculated using the DDH functional in a 216-atom supercell. \textbf{b.} Atomic displacements associated with the localized phonon modes. Only atoms near the defect are shown for clarity.}
    \label{fig:phonon_ipr}
\end{figure}

Supplementary Figure \ref{fig:phonon_ipr}a shows the inverse participation ratio of the computed phonon modes. We note that there are four localized modes: a two-fold degenerate mode at 13.81 meV, a non-degenerate mode at 15.99 meV, and another non-degenerate mode at 145.20 meV. Supplementary Figure \ref{fig:phonon_ipr}b shows the reference ground state structure and the displacements associated with the four localized phonon modes. We note that the modes are largely localized near the NV$^-$ defect. The mode at 13.81 meV is an $e$-mode, at 15.99 meV is an $a_1$-mode. The high energy mode at 145.20 meV corresponds to the N-O stretching mode (this value is similar to that of the stretching mode of the NO molecule, 232 meV \cite{laane1980characterization}).

While the spin-phonon coupling function would have to be calculated to identify which phonon modes couple with the NV$^-$ defect \cite{mondal2023spin}, given that the phonon energies generally exceed the zero-field splitting of this defect (46 GHz, i.e., 0.190 meV, at the DDH level), the spin relaxation time $T_1$ might be limited by the slower two-phonon processes as opposed to one-phonon direct and Orbach relaxation mechanisms. This could be favorable for the defect's $T_1$ value.

\clearpage
\section{Strain tests}
To identify which strain regime could be favorable in reducing the vibronic coupling, we carried out $\Delta$SCF calculations (total energy differences with different occupation numbers) at the PBE level for a 4x4x4 supercell, with 120 Ry plane-wave cutoff. Note that from the defect configuration shown in Fig.3a of the main text, $x$- and $y$-directions are equivalent. 

\begin{table}[H]
    \caption{\justifying Effect of uniaxial strain on the optical properties of the NV$^-$ center in MgO, calculated using $\Delta$SCF-PBE.}
    \centering
    \begin{tabular}{cccc}
    \hline
    \hline
        Property & 0\% & -1\% along x & -1\% along z \\
        \hline
        Absorption (eV) & 3.57 & 3.52 & 3.63 \\
        Emission (eV) & 1.55 & 1.59 & 1.57 \\
        ZPL (eV) & 2.45 & 2.50 & 2.48 \\
        $\Delta Q$ (amu$^{0.5}$\AA) & 3.37 & 3.15 & 3.47 \\
        \hline
        \hline
    \end{tabular}
    \label{tab:uniax}
\end{table}

For the uniaxial strain case, we first strained the unit cell by 1\% (compressive) along the $x$-direction and relaxed the structure by allowing the lattice parameters along $y$- and $z$-directions to relax. We then created a 4x4x4 supercell with the defect and  the new lattice parameters. Similarly, we strained the unit cell by 1\% (compressive) along the $z$-direction and relaxed the structure by allowing the lattice parameters along the $x$- and $y$-directions to be optimized, and then created a 4x4x4 supercell with the defect and the new lattice parameters.

\begin{table}[H]
    \caption{\justifying Effect of biaxial strain on the optical properties of the NV$^-$ center in MgO, calculated using $\Delta$SCF-PBE.}
    \centering
    \begin{tabular}{ccccc}
    \hline
    \hline
        Property & 0\% & -4\% along xz & -1\% along xy \\
        \hline
        Absorption (eV) & 3.57 & 3.60 & 3.47 \\
        Emission (eV) & 1.55 & 1.50 & 1.87 \\
        ZPL (eV) & 2.45 & 2.27 & 2.70 \\
        $\Delta Q$ (amu$^{0.5}$\AA) & 3.37 & 4.44 & 2.26 \\
        \hline
        \hline
    \end{tabular}
    \label{tab:biax}
\end{table}

For the biaxial strain case, we first compressively strained the unit cell by 4\% (1\%) along the $xz$ ($xy$) plane and optimized the structure by allowing the lattice parameters along the $y$- ($z$-) direction to vary. We then created a 4x4x4 supercell with the defect and the new lattice parameters.

\begin{figure}[H]
    \centering
    \includegraphics[width=0.75\columnwidth]{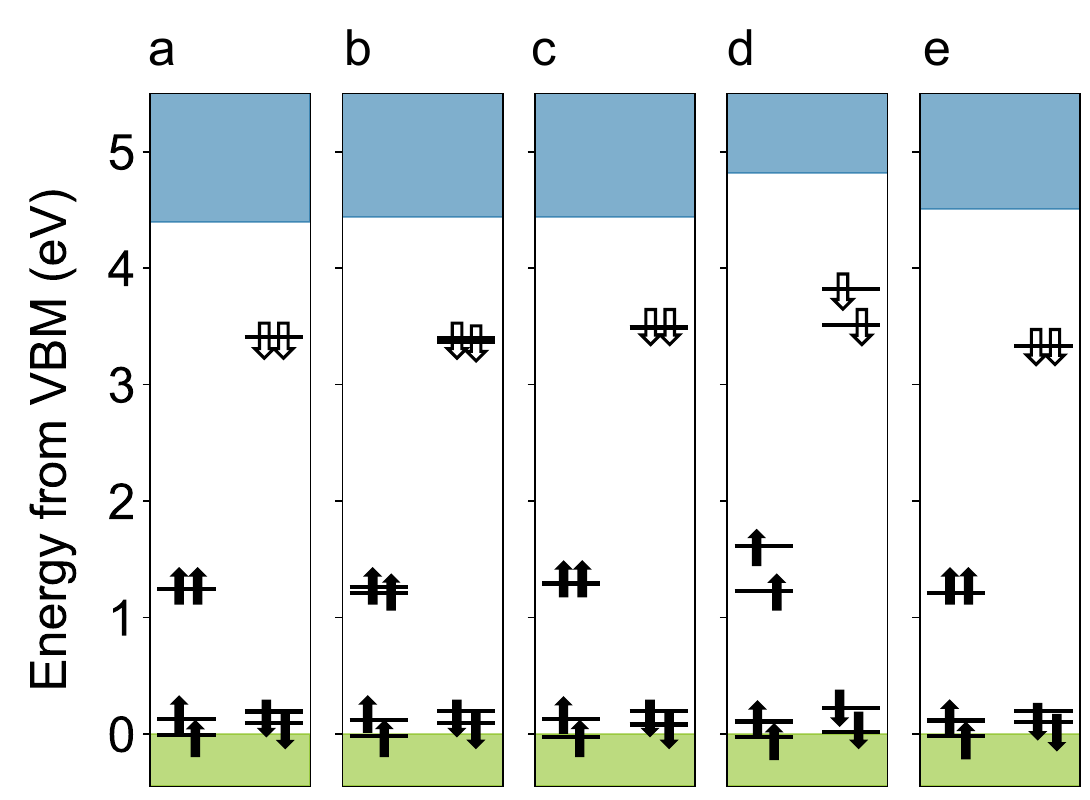}
    \caption{\justifying Defect level diagram of the NV$^-$ center in MgO under different strain cases, calculated at the PBE level. \textbf{(a)} unstrained \textbf{(b)} -1\% along x \textbf{(c)} -1\% along z \textbf{(d)} -4\% along xz \textbf{(e)} -1\% along xy}
    \label{fig:strain-dld}
\end{figure}

Note that we also considered the effect of tensile strain, but we faced convergence issues with the $\Delta$SCF method. However, it is unlikely that tensile strain will help reduce vibronic coupling. For example, in our test for 1\% biaxial tensile strain along the $xy$-plane, the partially optimized structure in the excited state  had a $\Delta Q$ of 4.44 amu$^{0.5}$\AA.

Supplementary Table \ref{tab:uniax} and Supplementary Table \ref{tab:biax} list the effect of these different strain configurations on the optical transitions energies of the NV$^-$ defect in MgO. Supplementary Figure \ref{fig:strain-dld} shows the effect of strain on the Kohn-Sham defect levels of the NV$^-$ center in MgO.
\clearpage
\section{Convergence tests for plane-wave cutoff energies}
We determined our plane-wave cutoff energy based on the convergence of the energy difference between the $e$-state in the spin up channel and the valence band maximum for the 216-atom supercell with the NV$^-$ defect. Calculations were done at the $\Gamma$-point with the DDH functional. As seen in Supplementary Figure \ref{fig:conv_test}, the energy difference is converged to 0.6 meV with a cutoff energy of 80 Ry, and to 7 meV when the cutoff energy is 60 Ry. Therefore, we use 80 Ry for our DFT-DDH calculations, and 60 Ry for the more expensive TDDFT-DDH calculations. Note that the small differences in energy (10$^{-4}$ eV) beyond a plane-wave cutoff of 90 Ry are due to differences in convergence of the self-consistency cycle. 

\begin{figure}[H]
    \centering
    \includegraphics[width=0.75\columnwidth]{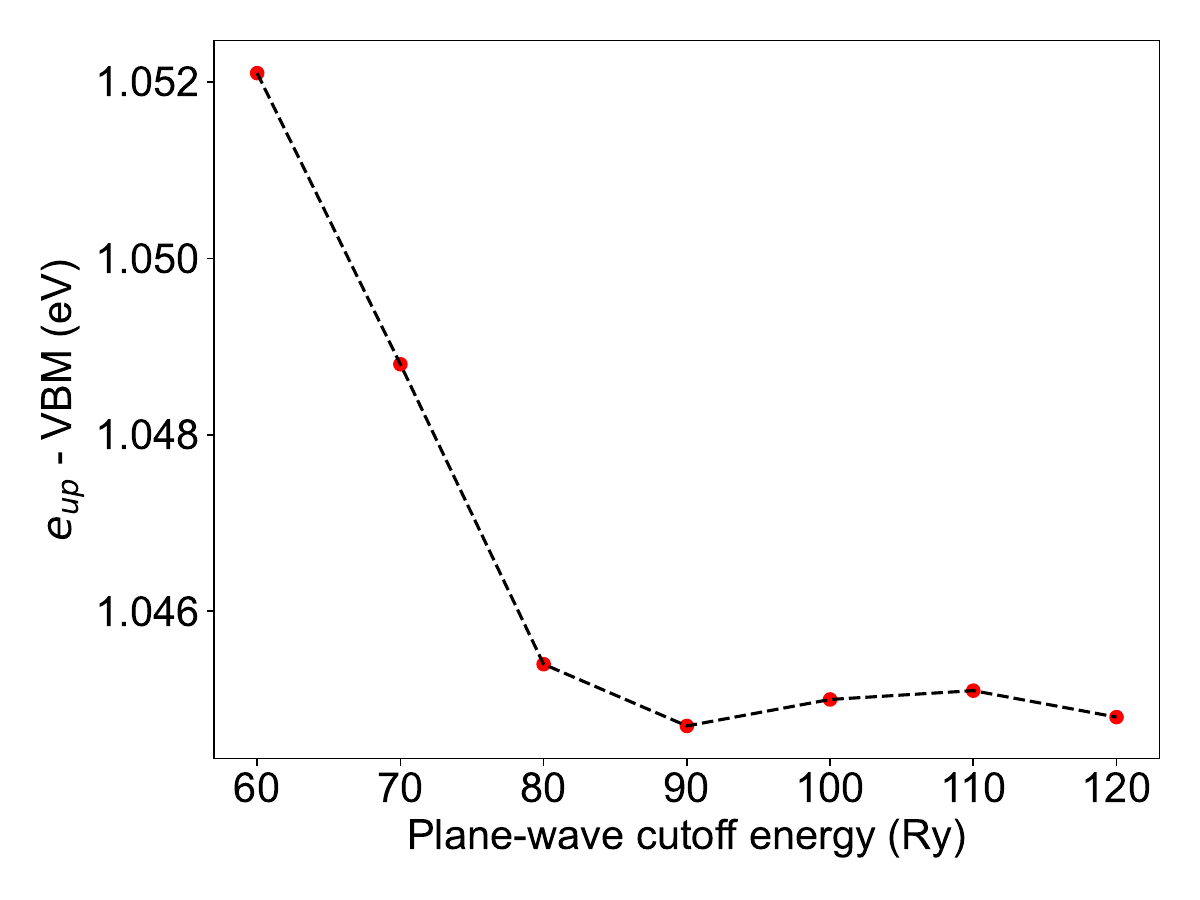}
    \caption{\justifying Energy difference between the $e$-state in the spin up channel and the valence band maximum as a function of plane-wave cutoff energy.}
    \label{fig:conv_test}
\end{figure}
\clearpage
\section{Experimental vs. DDH lattice constants}
We computed the equilibrium DDH lattice constant by fitting the total energies of the 8-atom unit cell of MgO at different lattice constants to the Murnaghan equation of state \cite{fu1983first}. A plane-wave cutoff of 80 Ry was used, and the Brillouin zone was sampled with a 6 x 6 x 6 k-point mesh. The fit is shown in Supplementary Figure \ref{fig:BM_fit}. We obtained the DDH lattice constant to be 4.195 \AA, in close agreement with the experimental value of 4.19 \AA.

\begin{figure}[H]
    \centering
    \includegraphics[width=0.75\columnwidth]{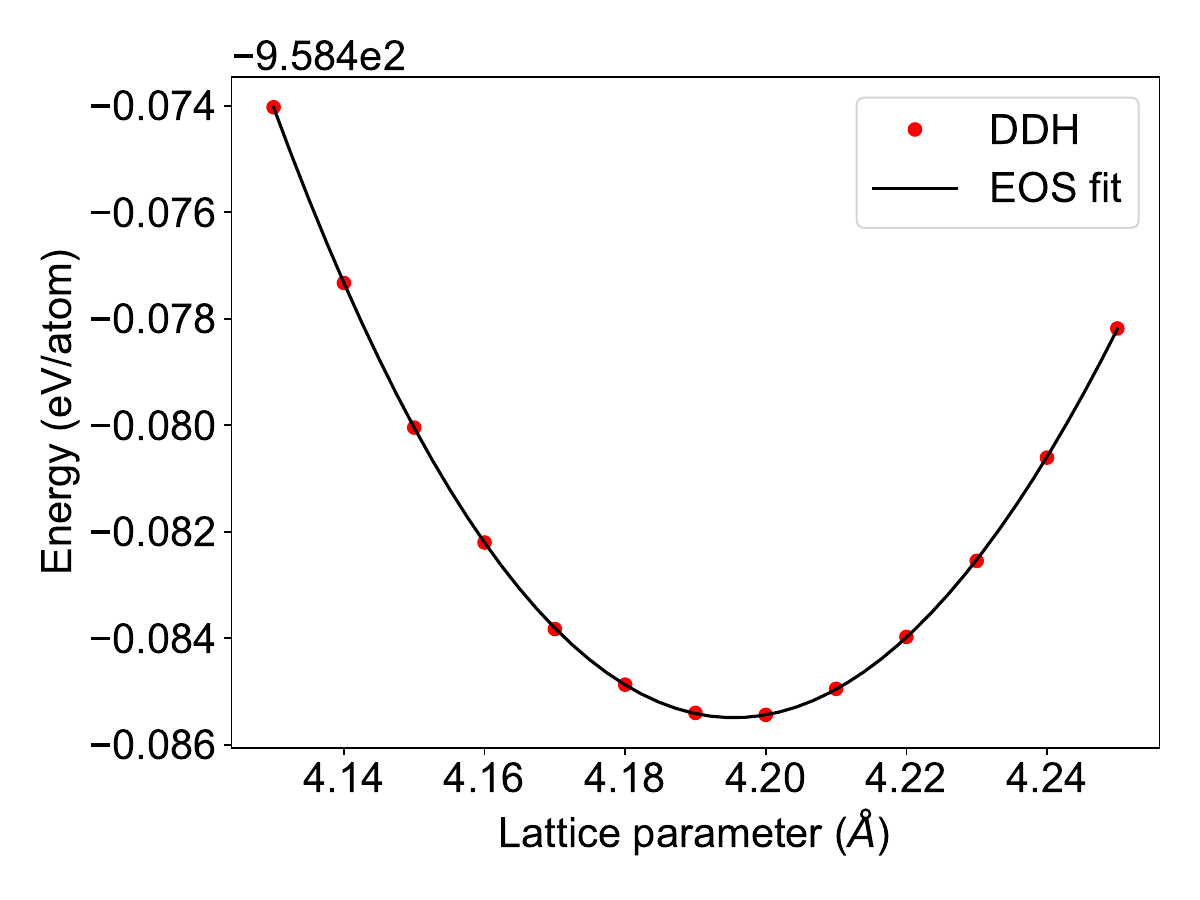}
    \caption{\justifying Total energy vs. lattice parameter for MgO unit cell and the corresponding equation of state fit.}
    \label{fig:BM_fit}
\end{figure}
\clearpage
\section{Calculation of O$_2$ and N$_2$ chemical potentials}
We used the HSE functional \cite{krukau2006influence} to obtain $E_{DFT, O_2}$ and $E_{DFT, N_2}$. Calculations of $E_{DFT, O_2}$ and $E_{DFT, N_2}$ were done in a cubic box of side length 12.57 \AA, plane-wave cutoff of 80 Ry, and with $\Gamma$-point sampling. As seen in Supplementary Table \ref{tab:O2_N2}, the HSE-computed binding energy and bond length of the respective molecules match most closely with experiments.

\begin{table}[H]
    \caption{\justifying Binding energy and bond length of O$_2$ and N$_2$ with different exchange correlation functionals.}
    \centering
    \begin{tabular}{ccccc}
    \hline
    \hline
        Quantity & DDH ($\alpha$=0.34) & HSE & PBE0 & Experiment \\
        \hline
        O$_2$ binding \\energy (eV) & -4.39 & -5.14 & -5.13 & -5.12 \cite{pople1989gaussian,allison_nist_janaf} \\
        O$_2$ bond \\length (\AA) & 1.186 & 1.194 & 1.194 & 1.208 \cite{huber1979constants} \\
        N$_2$ binding \\energy (eV) & -9.10 & -9.34 & -9.33 & -9.76 \cite{pople1989gaussian, allison_nist_janaf} \\
        N$_2$ bond \\length (\AA) & 1.085 & 1.084 & 1.089 & 1.098 \cite{huber1979constants} \\
        \hline
        \hline
    \end{tabular}
    \label{tab:O2_N2}
\end{table}
\clearpage
\section{Defect formation energies using PBE vs. DDH functional}
\begin{figure}[h]
    \centering
    \includegraphics[width=1\columnwidth]{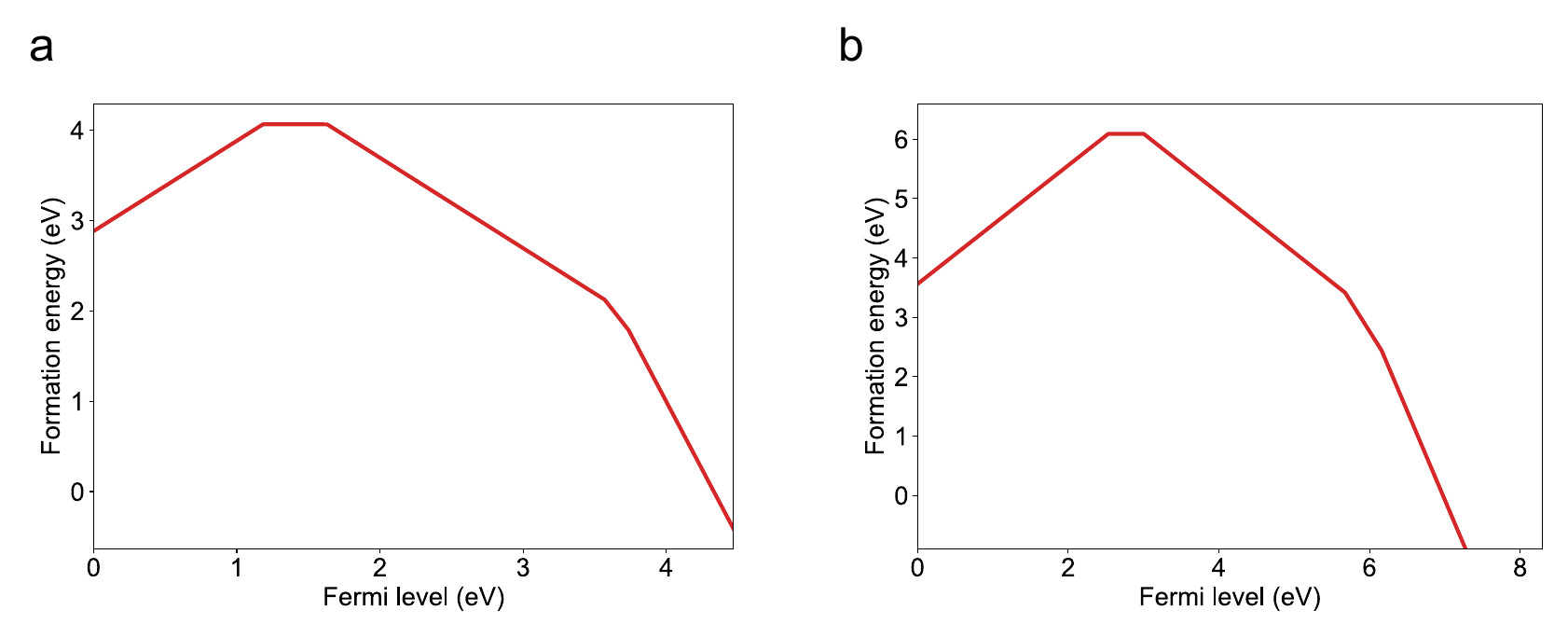}
    \caption{\justifying Defect formation energies of the NV defect in MgO as a function of Fermi level using \textbf{a.} PBE \textbf{b.} DDH exchange correlation functionals. The slope of the line corresponds to the charge state of the defect.}
    \label{fig:pbe_ddh_formnE}
\end{figure}

PBE is less accurate than hybrid exchange-correlation functionals for calculating defect formation energies. However, performing hybrid calculations for all the 3000 defects is computationally prohibitive. Moreover, while PBE systematically underestimates the formation energies, band gaps, etc., it still follows the general trends obtained using hybrid functionals. To check this, we compared the defect formation energies of the NV defect in MgO as a function of Fermi level at both the PBE and DDH level (Supplementary Figure \ref{fig:pbe_ddh_formnE}). We see that while the band gap and formation energies are underestimated with PBE (as expected), the relative widths of the stability regions of the different charge states are nearly identical. PBE also predicts the ground state of NV$^-$ to be a spin triplet, same as DDH. Thus, using PBE for screening purposes in this system (especially since there are no d- or f-elements involved), is valid.

\clearpage
\bibliography{si}